# High Spatiotemporal Resolution Structured Illumination Microscopy: Principle, Instrumentation, and Applications

Han Wang[1,2,†], Wenshu Wang[1,2,†], Xinzhu Xu[1,2], Meiqi Li[2,3,*] and Peng Xi[1,2,*]

[1] College of Future Technology, Peking University, Beijing, China, 100871;
[2] National Biomedical Imaging Center, Peking University, Beijing, China, 100871;
[3] School of Life Sciences, Peking University, Beijing, China, 100871;
† These co-authors contribute to this work equally.
* Email address: limeiqi@pku.edu.cn; xipeng@pku.edu.cn

## Abstract

Among super-resolution microscopy techniques, structured illumination microscopy (SIM) shows great advances of low phototoxicity, high speed, and excellent performance in long-term dynamic observation, making it especially suitable for live-cell imaging. This review delves into the principles, instrumentation, and applications of SIM, highlighting its capabilities in achieving high spatiotemporal resolution. Two types of structured illumination mechanics are employed: (1) stripe-based SIM, where the illumination stripes are formed through interference or projection, with extended resolution achieved through Fourier-domain extension; (2) point-scanning based SIM, where illumination pattern are generated through projection of the focal point or focal array, with extended resolution achieved through photon reassignment. We discuss the evolution of SIM from mechanical to high-speed photoelectric devices, such as spatial light modulators, digital micromirror devices, galvanometers, *etc.*, which significantly enhance imaging speed, resolution, and modulation flexibility. The review also explores SIM's applications in biological research, particularly in live-cell imaging and cellular interaction studies, providing insights into disease mechanisms and cellular functions. We conclude by outlining the future directions of SIM in life sciences. With the advancement of imaging techniques and reconstruction algorithms, SIM is poised to bring revolutionary impacts to frontier research fields, offering new avenues for exploring the intricacies of cellular biology.

Fluorescence microscopy plays an important role in studying the dynamics in life sciences, as the invisible cellular organelles can be specifically lighten up.[1-3] The resolution of conventional fluorescence microscopy techniques is limited by the wave nature of optics, known as the diffraction limit.[4,5] Super-resolution microscopy (SRM) breaks the long-standing diffraction limit in optical microscopy, making it possible to observe life processes and rather precise structures with a spatial resolution beyond the diffraction limitation in living specimens. Therefore, the 2014 Nobel Prize in Chemistry was awarded for the development of SRM.[6-10] The fast development and wide application of SRM has led to a surge in discovery of new life phenomena and a better understanding of the rules behind them.[11,12] Among all SRM techniques, structured illumination microscopy (SIM) stands out with its intrinsic advantages of low phototoxicity, compatibility with conventional fluorophores, multi-colour imaging, and high spatiotemporal resolution.[13,14]

As the name suggests, SIM refers to the modulation of illumination into a structured pattern, and the extraction of super-resolution (SR) information through pattern altering. Generally speaking, SIM can be classified into two categories. The first is achieved by directly generating a structured pattern on the sample plane, through interference[15], or projection[16], for instance. The second is achieved by multi-focus points scanning in the region of interest. Stripe-based SIM is the most classic implementation.[15] Conventional SIM implementations, mainly referring to 2D- and 3D-SIM, generate a sinusoidal structured illumination pattern with two- or three-beam interference.[15,17] For point-scanning SIM techniques, the patterned images are processed digitally or optically during the multi-focus scanning process to achieve super-resolution.[14,18] In both cases, multiple raw images are required to carry out super-resolution reconstruction. Therefore, the imaging speed is determined by the number of raw images to be taken, and the imaging quality relies on the accurate manipulation of the structured illumination patterns, such as directions and phases in stripe-based SIM and stepping in point-scanning SIM.

In the past two decades, the continuous demand for faster, deeper, clearer, and higher-dimensional imaging in the frontier of biological researches has brought comprehensive improvements to SIM.[18] The ultrastructure, dynamics, and interactions of the subcellular organelles in live cells expect microscopic techniques with both extended resolution and reduced artefacts, and therefore high spatiotemporal resolution SIM plays an indispensable role in 4D dynamic live-cell imaging.

In this review, we mainly focus on the principles, instrumentation, and applications of techniques achieving high spatiotemporal resolution SIM. In the first section, fundamental principles of super-resolution SIM (SR-SIM) is introduced, which can be categorized by the different mechanisms of structured illumination generation: 1) Stripe-based SIM, and 2) point-scanning based SIM. The second section reviews the instrumentation of advanced high spatiotemporal resolution SIM systems, focusing on the evolution from mechanical devices to high-speed photoelectric devices such as spatial light modulators (SLM), digital micromirror devices (DMD), and galvanometers (Galvo). These advancements have significantly improved the speed and resolution of SIM systems. The third section discusses the application of high spatiotemporal resolution SIM in biological research, presenting its outstanding live-cell SR imaging capabilities and pointing out some research fields where high spatiotemporal resolution SIM holds great potential. We conclude by outlining the future directions of SIM, including its integration with other imaging modalities like multiplane imaging, correlative light and electron microscopy, *Förster* resonance energy transfer (FRET), fluorescence life-time imaging microscopy (FLIM), label-free microscopy, and new dimensional imaging (time, spectrum, polarization), which promise to further expand the utility of SIM in life sciences.

## Principles

### Stripe-based SIM

In practical non-coherent optical systems, the image of an ideal point source, termed as the point spread function (PSF) of the system, is blurred to an Airy disk (Fig. 1a) due to diffraction. When two emitters are too close to each other, they cannot be distinguished. The resolution limit of optical microscopes was proved by Ernst Abbe in 1873 as

$$r = \frac{\lambda}{2\mathrm{NA}}$$

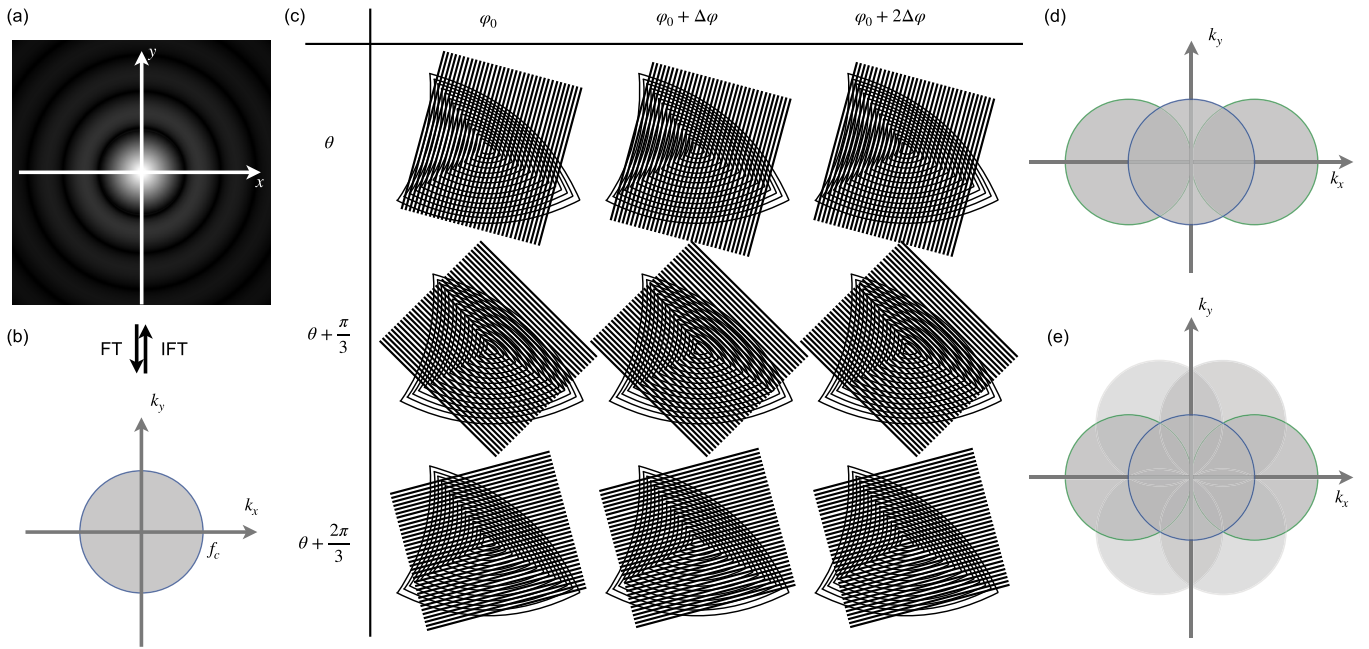

**Fig. 1 Basic principle of SR-SIM.** (a) The PSF of a microscope, which can be obtained from system calibration. The OTF and the PSF are a Fourier transform pair. FT: Fourier transform; IFT: inverse Fourier transform. (b) The OTF of a conventional wide-field microscope, which shows the observable region in the frequency domain. (c) Moiré patterns reveal detailed structural information. When two fringes superpose on each other, low frequency fringes will appear. Different angles of superposition lead to moiré fringes of different directions and frequency. As practically applied in SR-SIM, fringes in three directions (rows) are given. In each direction, fringes are phase shifted three times with a step of $\Delta\varphi$. A subtle phase shift leads to an obvious change in the low-frequency moiré fringes. (d) A horizontal sinusoidal pattern with a spatial frequency $f_c$ expands the observable region (the gray area). (e) Achieving isotropic resolution enhancement by rotating the structured illumination pattern.

where NA is the numerical aperture of the objective.[19] The minimum value of $r$, known as the famous diffraction limit, is approximately 200 nm when observing samples with visible light ($\lambda \geqslant 400$ nm) in air (NA $\leqslant 1$). The Fourier transform of PSF is termed as the optical transfer function (OTF) of the microscope, *i.e.*,

$$\text{OTF}(\boldsymbol{k}) = \mathscr{F}\left\{\text{PSF}(\boldsymbol{x})\right\}$$

where $\boldsymbol{x}$ is the spatial position vector, and $\boldsymbol{k}$ is the corresponding frequency vector.

From the perspective of signals and systems, any practical imaging system can be viewed as a low-pass filter, where the cut-off frequency of the OTF passband $f_c$ (Fig. 1b) is determined as:[15]

$$f_c = \frac{1}{r} = \frac{2\text{NA}}{\lambda}$$

In wide-field (WF) optical microscopy (Fig. 2a), the high-frequency components which contain detailed spatial information of the specimen are filtered out, leading to blurring due to inadequate resolution. The essence of SR-SIM lies right in expanding the observable region in the frequency domain when the cutoff frequency is limited as stated above. Moiré phenomenon provides a solution for such expansion: when patterns with certain spatial intensity distributions are superimposed on each other, the beat effect will generate low-frequency moiré fringes (Fig. 1c). Moiré fringes bring the high-frequency spatial information of the specimen into low-frequency region, allowing us to bypass the low-pass filtering nature of optical imaging. As practically applied in SR-SIM, fringes in three directions (as shown in different rows in Fig. 1c) are generated, which expand the cutoff frequency of the optical transfer function (OTF) at the corresponding directions. In each direction, fringes are phase shifted three times with a step of $\Delta\varphi$ (figures in the same row in Fig. 1c). From Fig. 1c, two intuitive conclusions can be achieved: 1) To reveal the information of the complete original pattern, illumination in different directions is necessary. 2) A subtle phase shift leads to a great change in the low-frequency moiré fringes, which means that the angle and phase information of the illumination pattern are critical in the reconstruction. Otherwise, it leads to artifacts. In the frequency domain, such low-frequency moiré fringes are able to pass through the low-pass optical system, and if it is possible to recover the high-frequency components of the original patterns, the observable region is essentially expanded and super-resolution is achieved. Fig. 1d gives an instance which shows how sinusoidal patterns with a spatial frequency $f_c$ expand the observable region (the gray area). To obtain isotropic resolution enhancement, the structured illumination pattern is typically rotated 3 times and 60° apart, as shown in Fig. 1c and 1e.

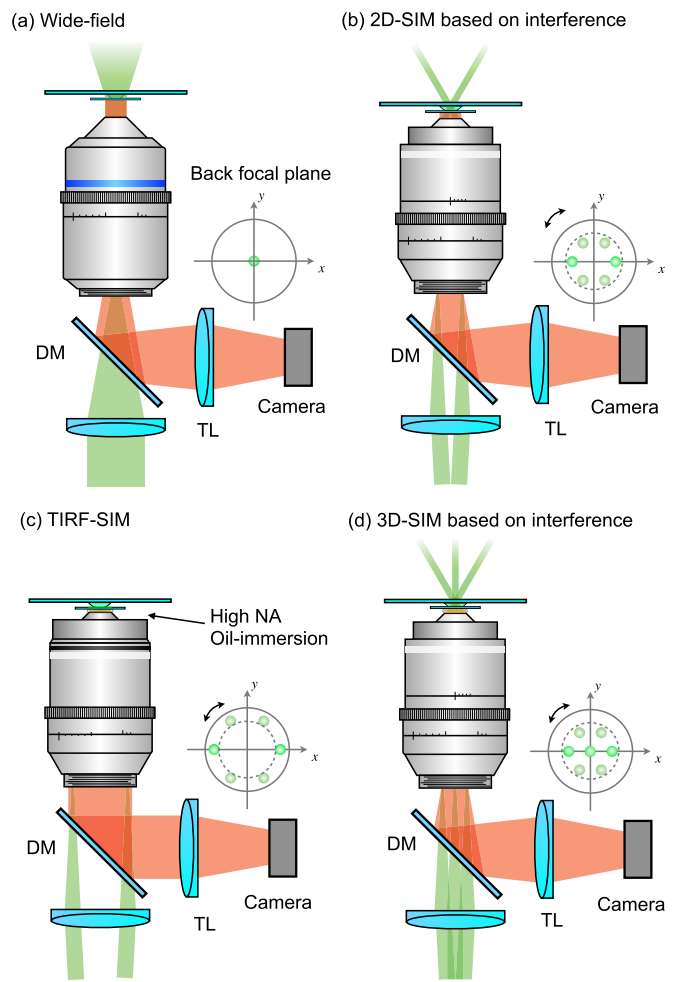

Fig. 2 Wide-field microscopy and interference-based SIM techniques. (a) Wide-field microscopy. (b) 2D-SIM based on interference. The laser beam pair should be rotated three times for isotropic resolution enhancement, as shown in the back focal plane (BFP) coordinate. (c) TIRF-SIM. In the BFP coordinate, the area outside the dashed circle is the TIRF region. Total internal reflection occurs only when incident light beams are outside the dashed circle. To make the TIRF region as large as possible, high NA oil-immersion objectives (1.40, 1.45, 1.49 or even 1.7) are commonly used. The interference pattern appears only *on* the bottom surface of the sample and attenuate exponentially as penetration depth grows, which is termed as evanescent wave.(d) 3D-SIM based on interference. The laser beam triplet should be rotated three times for isotropic resolution enhancement. DM: dichroic mirror; TL: tube lens.

In SR-SIM, the two *original patterns* stand for the spatial distribution of the specimen and the structured illumination, respectively. As the parameters of the illumination pattern, *e.g.*, spatial frequency and initial phase, are predefined and clearly known, the spatial distribution of the specimen can be obtained with the observed moiré fringes, and therefore, the diffraction limit is managed to be *evaded*. This process is essentially a process of solving linear equations. Taking 2D-SIM as an example (Fig. 2b), three linearly independent equations are needed to solve the ternary linear equations, the establishment of which comes from different initial phases $\varphi_1$, $\varphi_2$, and $\varphi_3$ of the illumination pattern. Therefore, phase shift is imperative in SIM implementations. Since the illumination pattern is also diffraction limited, the resolution of SR-SIM under linear fluorescence response is no higher than double the diffraction limit. When applying oil-immersion objectives and at total internal reflection fluorescence (TIRF) mode, the resolution of 85 nm can be achieved.[20]

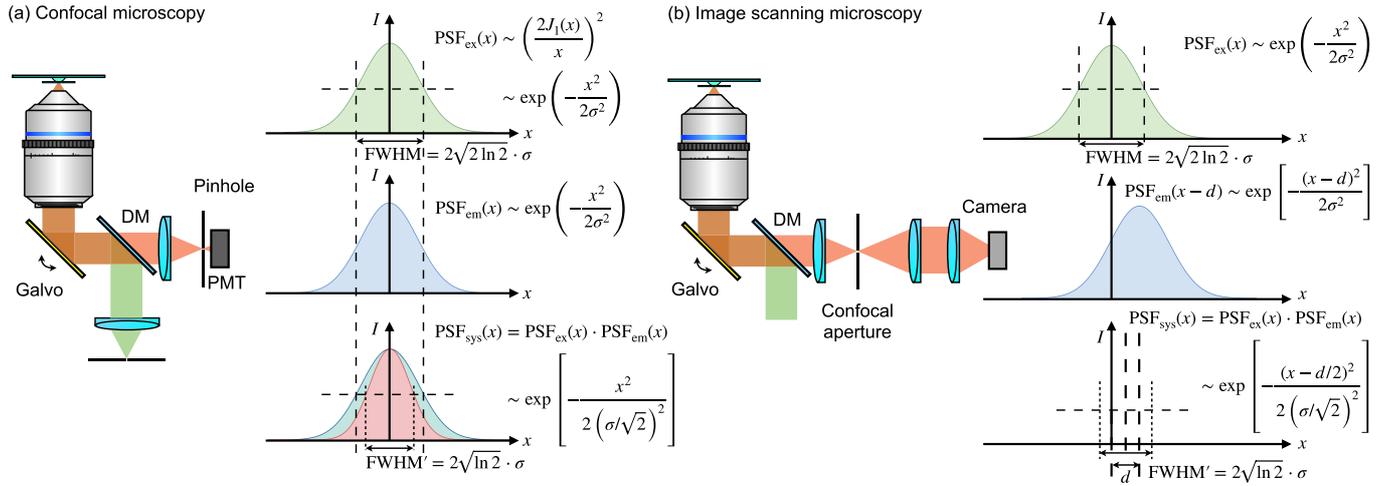

**Fig. 3 Confocal microscopy and image scanning microscopy.** (a) Confocal microscopy. Galvanometers are employed to scan the excitation beams (shown in green) on the sample plane and descan the emitted fluorescence (shown in red). In-focus fluorescence passes through the pinhole and is detected by the photomultiplier tube (PMT), while out-of-focus fluorescence are rejected. The product of the excitation $\mathrm{PSF}_{\mathrm{ex}}$ and the emission detection $\mathrm{PSF}_{\mathrm{em}}$ is the effective PSF of the microscope. Gaussian functions can be applied to approximate the profiles of PSFs. Theoretically, the FWHM of $\mathrm{PSF}_{\mathrm{sys}}$ can be compressed by a factor of $\sqrt{2}$. $J_1(x)$ is the first order Bessel function. (b) Image scanning microscopy. The confocal aperture eliminates background fluorescence. A camera is used for multi-point detection. For a detecting unit (pixel) that deviates from the central position with a distance $d$, the $\mathrm{PSF}_{\mathrm{em}}$ deviates correspondingly, making the product $\mathrm{PSF}_{\mathrm{sys}}$ deviate $d/2$ from the central position. Optical reassignment helps improve the SNR by correcting the deviation.

Over the past two decades, as the need for biological sample observation has grown, SIM has continuously evolved by integrating other optical approaches, resulting in numerous variants. Many of these variants have already found wide applications, such as:

1) In 2D-SIM, structured patterns can also be generated directly by projecting stripes onto the sample plane.[16] The difference between projection-based SIM and interference based SIM lies on the position of the pattern generator (such as SLM or DMD). For projection-based SIM, the pattern generator locates precisely on the conjugate plane of the imaging plane, so that the same frequency can be projected on the specimen. Consequently, the contrast of the pattern is subject to the modulation transfer function (MTF) of the objective (which is highly related with the objective NA), and higher frequencies are with lower contrast. For interference-based SIM, since the pattern is generated through Young's double slit interference, its contrast is not limited by the NA of the objective, but the polarization state.[21] And the position of the pattern generator need not to be precisely conjugated to the imaging plane.

2) In TIRF-SIM, the excitation frequency is higher than the cutoff frequency of the detection OTF (dashed circle in Fig. 2c)[22-24];. Therefore, the spatial resolution can be improved beyond two folds.

3) In 3D-SIM, three-beam interference are typically employed, and 15 raw images (3 angles × 5 phases) are typically collected.

4) In nonlinear-SIM (NL-SIM), a further expansion of the OTF is achieved through nonlinear effects such as photoswitching and saturation, and multiple angles and phases should be collected correspondingly[25,26].

5) In speckle SIM, random speckle patterns are employed, followed by blind-SIM, which is insensitive to the initial angle or phase of the structured pattern[27,28].

It is important to note that, the super-resolution capability of SIM is fundamentally dependent on the accurate generation of the illumination pattern onto the sample plane. However, this condition can be disrupted by aberrations and scattering within cells. Since SIM reconstruction is typically performed in the frequency domain, such disruptions can lead to frequency-domain errors, resulting in various artifacts in the reconstructed images.[29] In essence, if the structured illumination condition is compromised, the SR-SIM result will be impaired.

**Point-scanning based SIM**

In a broader context, another form of structured illumination encompasses focal spot or focal array illumination. Confocal microscopy, which employs point-scanning or multiple focal array scanning—such as in spinning disk confocal—is widely utilized in biological imaging of thick specimen to counteract aberrations and scattering.[30] The emergence of point-scanning based SR-SIM systems has been on the rise for this very reason.

The imaging process is essentially the convolution between sample and the PSF: $d(x) = s(x) \otimes \text{PSF}(x)$. In confocal microscopy the sample is excited with a focused light spot, and only the fluorescence emitted from the central part of the focused excitation light spot can be received and recorded, thereby effectively separating in-focus light from out-of-focus background. By introducing a pair of conjugate pinholes into the illumination system (Fig. 3a), we have the excitation PSF and the fluorescence emission detection PSF, denoted as $\text{PSF}_{\text{ex}}$ and $\text{PSF}_{\text{em}}$, respectively. Theoretically, they share the same profile that can be approximated with a Gaussian function. The product of $\text{PSF}_{\text{ex}}$ and $\text{PSF}_{\text{em}}$ is the effective PSF of the system:

$$\text{PSF}_{\text{sys}}(x) = \text{PSF}_{\text{ex}}(x) \cdot \text{PSF}_{\text{em}}(x).$$

As a result, confocal microscopy effectively *compresses* the system's PSF by a factor of $\sqrt{2}$ at maximum thanks to the PSF multiplication, improving the spatial resolution. In practice, the pinholes cannot be infinitely small, since a smaller pinhole leads to a significant decrease in SNR. However, enlarging the pinhole for better SNR in turn leads to an expansion of $\text{PSF}_{\text{em}}$, compromising the confocal resolution. Therefore, there is actually only marginal lateral resolution improvement in practical confocal microscopy.[31]

To address the contradiction between resolution improvement and SNR, Colin Sheppard proposed a *theoretical* solution in 1988[32]: replacing the single-pixel detector with a multi-pixel detector array, where each pixel acts as a small pinhole detector, and post-processing techniques would correct the offset confocal images, thus improving resolution while fully utilizing the light signal. Over 20 years later, Müller *et al. experimentally* realized this concept, naming it image scanning microscopy (ISM), as illustrated in Fig. 3b.[33,34] In point-scanning based SIM, super-resolution is achieved through the imaging of confocal pinholes, following with shrinking the image by half, before summation to the next position. For the pixel in the central position conjugate to the excitation pinhole, it experiences a similar process to confocal microscopy. For other pixels that deviates from the central position with a distance $d$, their detection PSF also deviates with $d$. Therefore, the product between $\text{PSF}_{\text{ex}}(x)$ and $\text{PSF}_{\text{em}}(x-d)$ leads to a deviated overall PSF with a deviation $d/2$. With proper algorithms such deviation can be removed, and thus ISM takes full advantage of the PSF compression while keeping high SNR. ISM has a resolution of ~150 nm, approximately $\sqrt{2}$ times that of conventional wide-field or confocal laser scanning microscopes.[31] The resolution of ISM is similar to that of projection-based stripe SIM, because in both cases, the excitation PSF is limited by the excitation NA of the objective. Additionally, one may apply deconvolution to improve the resolution, which is outside the scope of this review.

**SR-SIM for three-dimensional imaging**

In addition to extending the lateral resolution, the comprehensive and 3D observation of biological samples is also a necessity for biologists. 3D optical microscopy has thus been rapidly advancing on the basis of 2D microscopy, and 3D-SIM stands out among various 3D SRM techniques due to its prominent advantages of 3D tomography, low phototoxicity, fast live-cell imaging, and great compatibility with various fluorescent labels. This section will focus on how SR-SIM works for imaging in higher dimensions.

3D-SIM inherits the overall layout and the advantages of 2D-SIM. Meanwhile, it expands 2D-SIM axially by extending the depth of field (DOF) and eliminating out-of-focus background. Generally speaking, 3D-SIM techniques can be separated into two categories, featuring *3D stripe modulation*, and *point-scanning 3D imaging*. Essentially, they all generate structured illumination patterns on the sample and thus obtain high-frequency components that carries sample details for super-resolution imaging. In contrast, 3D stripe modulation often requires the interference of multiple beams laterally and axially, to generate a structured pattern in the $xyz$ plane. *Point scanning-based 3D imaging* acquires information on one or multiple focal points, and the detection of the high-frequency requires information from adjacent pixels. It should be noted that, the 3D SR-SIM techniques are down-compatible to 2D imaging, *i.e.* both stripe modulation 3D-SIM and the point-scanning 3D-SIM can be performed for single slice (one 2D image) only.

*3D-SIM based on interference*

We have discussed how the sample modulated by structured patterns in 2D-SIM enables high-frequency information to pass through the OTF. In 2008, Gustafsson *et al.* first proposed to modulate the sample with 3D structured illumination on their theoretical and practical basis in 2D-SIM[15], thereby achieving super-resolution both laterally and axially.[17] They proved that 3D super-resolved information could also be extracted through structured illumination. Similar to the principles of 2D-SIM, the basic idea of 3D-SIM is also to expand the observable region so as to have more high-frequency components that carry fine spatial information details pass through the OTF. What is different from the 2D situation is that when observing the OTF of a wide-field microscope in the 3D frequency domain, the OTF exhibits a *torus* shape, indicating that the low-frequency components out of the focal plane cannot be received by the detector. This is known as the *missing-cone problem*, and therefore, it is the core mission to fill the missing region through 3D spectra shifting.

Three-beam interference is a typical method for 3D spectra shifting in 3D-SIM (Fig. 2d, Fig. 4a).[17] As shown in Fig. 4c, in the $xOy$ plane, the three tori have expanded the observable region, which is the result of the interference from the $\pm 1$-order coherent light beams, just as in 2D-SIM situations. In the 3D frequency domain, three pairs of missing cones are left, and they are filled by the interference from the zero-order beam with $\pm 1$-order beams (Fig. 4a). Similarly, for isotropic lateral resolution enhancement, the illumination pattern should also be rotated three times, and meanwhile, the rotation has well filled the new missing cones, improving the axial resolution as well (Fig. 4a). Following the principles above, it can be seen that through modulation, rotation, and phase shifting of the 3D structured illumination, the system's OTF is extended both laterally and axially. Once the parameters are determined, in many SIM reconstruction algorithms, a generalized Wiener filter can be applied to the components containing different parts of the spectrum and obtain the super-resolution reconstruction estimate of the sample's true spectrum.[35] The theoretical maximum expansion factor is $1 + \lambda_{em}/\lambda_{ex}$, where $\lambda_{em}$ is the emission wavelength, and $\lambda_{ex}$ represents the excitation wavelength.

Note that since the objective can only collect back-propagating light, axial resolution is typically inferior to lateral resolution. 3D imaging of thicker samples still relies on axial scanning layer-by-layer. SIM based on 3D-PSF modulation achieves resolution enhancement in the axial direction, and that is its greatest advantage over pure slice-stack methods. To further enhance the axial resolution of 3D-SIM, Manton *et al.* in 2020 introduced a fourth axial light beam into 3D-SIM structured illumination interference using a mirror and a low-magnification objective placed behind the sample (Fig. 4b).[36] This approach further extends the system's OTF cutoff in the axial direction (Fig. 4c), achieving isotropic 3D super-resolution imaging with 125 nm resolution in both lateral and axial dimensions. Similarly, Li *et al.* uses a mirror to reflect the 0th order, and both lateral and axial resolution can also be improved to ~120 nm isotropically in their four-beam 3D-SIM (Fig. 4d left).[37] One step further, when employing the opposing-objective-lens geometry, all the three beams can be collected and reflected, enabling a six-beam interference and achieving an isotropic resolution of ~100 nm (Fig. 4d right).[38]

### 3D-SIM based on point-scanning

Laser scanning confocal microscopy (LSCM) is a classic implementation of point-scanning technology.[39] In LSCM, the region excited by the focused light spot is the excitation PSF, which takes an ellipsoid shape elongating axially. 3D information $(x, y, z)$ of the sample can be obtained through the relative movement of the excitation and the specimen. Galvanometers can be applied to achieve lateral scanning of the light spot across the sample. The sample stage or the objective lens is then moved axially to carry out axial scanning through the sample. Furthermore, LSCM can also take time and wavelength of the excitation light into consideration,

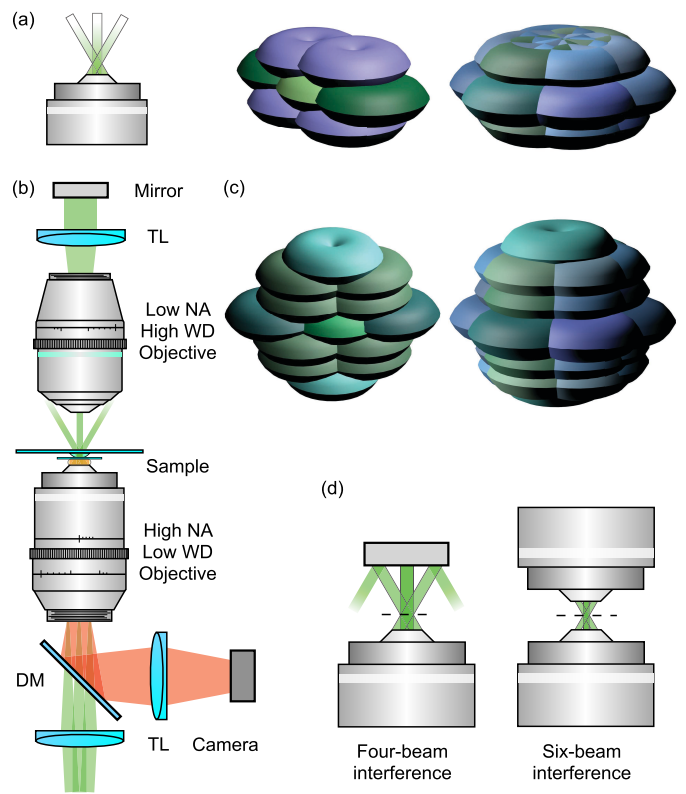

**Fig. 4 3D-SIM expands the observable area in the three-dimensional frequency domain.** (a) Three-beam interference (left), the spectrum in 3D frequency domain (middle) and the spectrum after pattern rotation (right). (a) Introducing a fourth axial beam into interference. A low NA, high working distance (WD) objective is employed to collect the zeroth beam. The mirror behind reflects the beam back to the sample plane and thus introducing a fourth axial light beam into interference. DM: dichroic mirror; TL: tube lens. (c) The observable region in 3D frequency domain. Full regions and their cross-section profiles are presented, which show the resolution improvement in the axial direction. (d) Simplified method for four-beam interference (left) and six beam interference (right).

which makes it possible for *4D imaging* $(x, y, z, t)$ and even *5D imaging* $(x, y, z, t, \lambda)$.[40]

ISM is a 3D confocal scanning method that well balances the pinhole size and signal strengths. One typical example of ISM is Zeiss Airyscan, which employs 32 independent detection units in a hexagonal array.[31] Each unit can be equivalently viewed as a confocal pinhole with a slightly different view on the sample. After the excited fluorescence spot is illuminated on the detection array, the signals from all detection units are appropriately weighted and the frequency components are assigned to the correct positions through linear deconvolution. This approach effectively provides the high resolution from an equivalent small pinhole while maintaining high light utilization efficiency and SNR.[41] Given that axial information is considered during the linear deconvolution step, Airyscan enhances both axial and lateral resolution, while maintaining high-SNR optical sectioning capabilities.[42] Similarly, using 25 square-tiled independent detection units, Nikon Inc. introduced the Nikon SPatial ARray Confocal system (NSPARC).[43] Each detection unit has a diameter of about 0.2 Airy units (AU), and the entire detection array has a diameter of about 1 AU, enabling oversampling of the emission light from a single Airy spot in the confocal plane. With the support of resolution enhancement algorithms based on ISM, its spatial resolution achieves 100 nm laterally and 300 nm axially.

To reconstruct a complete 3D structure, traditional ISM requires extensive axial scanning, which significantly reduces the system's 3D imaging speed. To improve the temporal resolution of ISM, researchers have combined ISM with PSF engineering, substantially extending the imaging depth of a single layer scan and enabling 3D reconstruction from 2D scanning.[44,45] A representative implementation of 3D-ISM is RESCH (REfocusing after SCanning using Helical phase engineering)[46], where helical phase modulation is introduced into the optical detection path, transforming the standard PSF into a Double Helix PSF (DH-PSF). The main advantage of RESCH is its ability to collect information from axial planes adjacent to the focal plane (with an axial range of approximately 400 nm for NA 1.35) while maintaining axial resolution, thereby reducing the overall image acquisition time and minimizing phototoxicity. However, the shape of the PSF of RESCH is broad in the y direction due to the anisotropic shapes of the two lobes.[46] For this reason, Roider *et al.* applied single-helix phase engineering and multi-view convolutional algorithms and proposed MD-RESCH (multi-view deconvolved RESCH) in 2015, which improves lateral resolution by approximately 20% compared to confocal microscopy, while maintaining axial resolution comparable to traditional ISM.[44] In 2016, Roider *et al.* further refined this technique, introducing a more general method known as engineered Image Scanning Microscopy (eISM). By sculpting the excitation and detection PSFs into helices with opposite handedness, eISM enables the capture of sample information in a volume extending over 4 times the z-resolution, with physical refocusing achievable within an axial range of 3 μm.[45] In addition, DH-PSF can also be combined with multifocal SIM (MSIM) to obtain MSIMH (MSIM with Helical phase engineering). In 2018, Li *et al.* achieved almost twice the lateral resolution of traditional widefield microscopy, with an extendable imaging depth range of up to 5.5 μm. Moreover, it enables faster acquisition than traditional RESCH, as MSIMH is able to complete a 30 μm × 30 μm field of view(FOV) acquisition in 6 seconds.[47]

## Instrumentation

The precise control of the illumination pattern is the key for SIM implementation. Instrumentation of various high spatiotemporal resolution SIM techniques will be introduced in this section. In stripe-based SIM, interference is an effective way to generate continuous periodic sinusoidal patterns, and is thus widely applied in SIM. Early SIM systems use diffraction gratings as the core element, after which the $\pm 1$-order diffraction light beams are coherent with each other. When they are superimposed on the sample plane, stripe patterns will appear.

Conventional stripe-based SIM is dependent on the rotation and shifting of structured illumination patterns. In systems where gratings are used, rotation and phase shifting depend on precise mechanical movements of the grating. Consequently, the imaging speed is restricted by these mechanical motions between capturing raw images. Additionally, gratings with different grating constants are required in such systems to match the resolution capabilities of objectives with different magnification. This necessitates manual replacement of gratings when switching objectives, significantly reducing convenience and operational flexibility, as seen in early commercial designs like Nikon's N-SIM E system.[48] Since multiple raw images are required for SR image reconstruction, it typically takes several seconds to acquire one SR image, which hinders its application in dynamic live-cell imaging. Furthermore, since it relies on high-contrast fringes with a relatively shallow penetration depth in biological samples, it is not well-suited for thick sample imaging. Similarly, in point-scanning SIM, the imaging speed is largely limited by the aspect ratio of the excitation PSF and the spatial period of the illumination pattern, that is, the number of raw images that need to be taken, as well as the precise stepping movement of the focal point(s).[30,49]

High spatiotemporal resolution SIM techniques demand high-speed photoelectric devices such as modulators and detectors, like spatial light modulators (SLM)[50-52], digital micromirror devices (DMD)[53-55], galvanometers (galvo)[56], and sensitive detectors such as scientific CMOS camera or detector array[57-60]. The speed limit of a SIM system is affected by multiple factors, including the illumination modulation, the detection sensitivity, the brightness of the fluorescence sample, and the photobleaching/phototoxicity issue, *etc.*. In this section, the instrumentation of different SIM techniques will be introduced and the key parameters including the spatial and temporal resolution will be listed in Table 1 in the end of this section.

**Stripe-based SIM**

Devices for structured illumination generation are crucial for SIM, and their switching speed is either determining or closely related to the imaging speed of the entire system. Similar to 2D-SIM, Gustafsson *et al.* first implemented 3D-SIM in 2008 with a fused silica linear transmission phase grating on a piezoelectric translation stage and an electric rotary stage, reaching lateral and axial resolutions of 103.9 nm and 279 nm, respectively.[17] With a need of 15 images per layer, 100 ms exposure per image, it takes 140 s for data acquisition for 80 layers of images with 512×512 pixels. The imaging speed is primarily limited by the mechanical movement of the phase grating and the exposure time of the EMCCD camera.

*Liquid crystal modulators*

To circumvent the inherent limitations of mechanical gratings, a family of liquid crystal (LC) devices, such as LC SLM and LC polarization modulators, is applied in SIM systems. SLM can generate pixelated grating patterns, allowing for rapid switching of direction and phase simply by writing new digital pattern data to the SLM. In 2008, Fiolka *et al.* applied SLM to SIM systems, and successfully constructed an SLM-based TIRF-SIM system that achieved a lateral resolution of 91 nm.[61] In 2009, Kner *et al.* further developed the TIRF-SIM system using a silicon-based ferroelectric LC SLM, which enabled clear observation of microtubule growth, shortening, and movement within cells at a video frame rate of 11 Hz over an 8×8 µm² field of view.[62] In the same year, Chang *et al.* performed phase changes in four directions with an SLM of a maximum refresh rate of 60 Hz, completing the acquisition of 12 images within 30 seconds.[63] In 2011, Shao *et al.* used a ferroelectric LC SLM and a series of phase retarders as coherent light generation and polarization modulation devices in a 3D-SIM system.[64] Compared to physical gratings, the switching speed is significantly increased. The researchers achieved lateral and axial resolutions of 120 nm and 360 nm, respectively, with exposure time reduced to 35 ms per layer.[64] The readout frequency of the EMCCD with a 14-bit analog-to-digital convertor is 10 MHz, and they reached a time resolution of 5 to 25 seconds in imaging microtubules of *Drosophila* S2 cells and mitochondria in HeLa cells. In 2012, Fiolka *et al.* employed a fast programmable LC polarization rotator and an sCMOS camera in their system, further reducing the exposure time to 5 ms for every single image, which significantly accelerated data acquisition, making it possible to complete 3D imaging of a 1 µm thick, 51 µm × 51 µm sample in just 1.4 s.[65] When observing samples of weak fluorescent emission, in order to maintain spatial resolution, the single-image exposure time was set to 20 ms. When imaging HeLa cells of 1.25 µm thick, the system completed volumetric imaging in 4 s in single-color mode and 8.5 s in dual-color mode. The four years from 2008 to 2012 saw the great spatiotemporal resolution improvement in SIM, and more advanced devices and techniques has since then been pushing SIM further. In 2014, Förster *et al.* designed the Fast-SIM system using an SLM with a refresh rate of approximately 1 kHz.[66] They employed a segmented polarization wave plate to ensure that the diffracted light in each direction was circularly polarized, enhancing both imaging quality and speed. In 2023, Li *et al.* achieved higher resolutions of 123.5 nm laterally and 163 nm axially by simply adding a mirror above the three interfering beams and achieved a four-beam interference method (Fig. 4d).[37] In comparison to the four-beam interference strategy illustrated in Fig. 4d, which was not widely applied due to its complicated optical setup, Li's method is far easier in implementation because of its minimal modification to the 3D-SIM system setup. In such system, the average acquisition time for each raw image is 26 ms, and 3D imaging of a 90 µm × 76 µm × 4 µm cell sample with an axial scanning step of 0.06 µm takes only 26.5 s. Moreover, more interfering beams can also be taken into consideration for 3D-SIM implementation.[38]

With the application of high-speed LC devices and more 2D- and 3D-SIM strategies, temporal resolution of SIM systems is primarily limited by the exposure time of the sCMOS camera rather than the time for structured illumination modulation. As camera imaging speeds gradually improve, SLM-based SIM also see a significant speed increase. Lu-Walther and Song increased the acquisition speed to 20.13 Hz (field of view size of 8.3×8.3 µm²) in 2015 and 79 Hz (field of view size of 16.5×16.5 µm²) in 2016.[67,68]

The combination of high-speed modulation devices and reconstruction algorithms significantly enhance SIM's temporal resolution. In 2018, Huang *et al.* proposed the Hessian algorithm, which takes full advantage of the continuity of biological structures across multiple dimensions as prior knowledge to guide reconstruction.[69] This algorithm can match sub-millisecond excitation pulses, achieving an imaging rate of 188 Hz in an SLM-based TIRF-SIM system. With Hessian-SIM, the system reached a raw data acquisition frame rate of 1692 Hz, corresponding to an SLM fringe exposure time of 590 µs. The trade-off for such high-speed imaging is the reduced frame size permissible by the camera, because reducing the number of active rows on the sCMOS sensor effectively shortens the camera's exposure-readout time. The system frame rate is ultimately constrained by the SLM fringe switching frequency. At the highest acquisition frame rate, the reconstruction frame size in Hessian-SIM was only 512×144 pixels.

Addressing the challenge of further increasing frame size and enhancing spatiotemporal information throughput (frame size × frame rate) at the extreme operating frequencies of high-speed fringe modulators, Peng Xi group presented a solution in their parallel acquisition-readout SIM (PAR-SIM) in 2024 (Fig. 5a).[70] The authors identified that the limitation on the size of each frame is due to the *rolling shutter readout process* of sCMOS sensors, where readout time is wasted. By employing highly efficient and precise synchronization signals, PAR-SIM cleverly utilizes the rolling shutter exposure-readout process of sCMOS sensors. Coupled with the SLM, which exposes different phases and directions of fringes in each sub-frame, they achieved direct output of 6 raw sub-images per frame (Fig. 5b), with a spatiotemporal information throughput of 133 Megapixels per second (Mpx/s) at 1352 × 384 pixels × 256 Hz. This throughput is approximately an order of magnitude higher than that of Hessian-SIM (512×144 pixels × 188 Hz) at 13.86 Mpx/s. At the same time, a resolution of 100 nm is obtained. The combination of high throughput and high resolution makes PAR-SIM a powerful tool for applications requiring high spatiotemporal resolution, high-resolution imaging. The specific principle is illustrated in Fig. 5b, showing how PAR-SIM exploits the *ramp* time difference in rolling shutter exposure.

*Digital micromirror device*

As imaging speed increases further, even the switching speed of LC SLMs cannot satisfy the need for high-speed imaging — there is an urgent need for a faster structured light modulator in SIM systems. In 2013, Dan *et al.* introduced a novel high speed modulation device, the DMD, into optical sectioning SIM (OS-SIM).[16] Compared to SLM, DMD has a significantly higher switching rate of nearly 32 kHz and can maintain stable patterns for several seconds without requiring a refresh cycle, which greatly reduces the complexity of temporal control. Moreover, DMDs are significantly more cost-effective, with costs ranging from 10% to 50% of SLMs, making them a more attractive option in terms of cost effectiveness.

DMD-based SIM systems can be categorized into two main types: the projection-based SIM proposed by Dan *et al.* in 2013[16] and the interference-based SIM introduced by Li *et al.* in 2020[71]. The projection-based SIM uses a low-coherence LED for illumination, offering the advantages of a simple structure and easy implementation of multi-color switching, while the low contrast of the projected patterns limits the system's resolution enhancement (Fig. 6a, b). In contrast, the interference-based SIM, which uses a high-coherence laser as the light source, can achieve a two-fold resolution improvement (Fig. 6c, d). However, due to the blazed grating effect, the diffractive orders and diffraction efficiency changes with respect to the corresponding incident wavelengths in interference-based SIM. To address this, Li *et al.* have simultaneously proposed the concept of multi-color DMD-SIM imaging. By adopting the ON and OFF states of the DMD at different incident angles to meet the illumination condition, DMD-SIM has the potential to perform four-color imaging. In 2021, Sandmeyer *et al.* constructed an interference-based SIM using DMD, achieving live-cell imaging at a video rate of up to 60 Hz over an 18×18 μm² field of view, allowing for the observation of dynamic changes in mitochondria and the endoplasmic reticulum (ER).[72] To better utilize multi-color information, Brown *et al.* developed a theoretical framework for coherent SIM using a DMD as a multi-color diffractive optical element in the same year.[73] Following this, Lachetta *et al.* developed a dual-color DMD-SIM system using temperature-controlled wavelength matching.[74] By adjusting the wavelengths of two lasers to match the grating's diffraction period, mechanical alignment was not required, achieving imaging rates of 60 Hz for single-color and 30 Hz for dual-color acquisition. To further overcome the dispersion caused by the DMD grating structure for different wavelengths, Gong *et al.* in 2023 used a grating to compensate for the DMD dispersion, achieving four-channel multi-color SIM imaging.[75] However, the suboptimal blazed conditions of the beam on the grating and DMD confront this method by excitation beam energy attenuation, posing new challenges to its application. In 2023, Li *et al.* from Peng Xi's Group proposed a high-speed 3D-SIM system utilizing an electro-optical modulator and a DMD, which has both polarization maintenance characteristics and polarization modulation capabilities, attaining a lateral resolution of 133 nm and an axial resolution of 300 nm, with a time resolution of 6.75 s for a 15 ms exposure time.[76] They verified the system's 3D SR imaging ability, high temporal resolution, and polarization imaging capabilities from experiments in imaging nuclear pore complexes, microtubules, actin filaments, intracellular mitochondria, plant and animal tissues, *etc.* The integration of polarization control in the 3D-SIM system also allowed for the exploration of anisotropic structures within the samples, providing additional contrast mechanisms that are not available in conventional SIM setups.

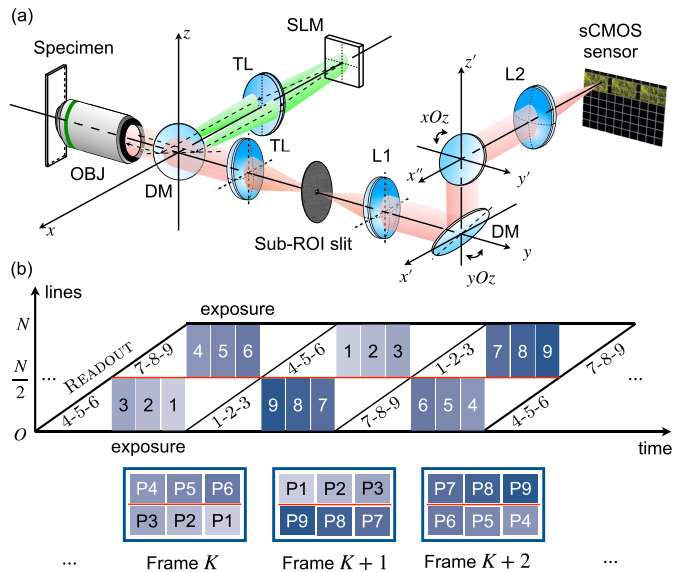

**Fig. 5 PAR-SIM setup and its principle.** (a) Optical path, showing the fringe excitation and detection parts of the PAR-SIM system composed of precise synchronization of the SLM, camera, and galvo mirror. (b) Utilizing the *ramp* time difference in the rolling shutter mode of the sCMOS, sub-ROIs in the respective 0-N/2 rows and N/2-N rows of the same chip regions are simultaneously exposed or read out. This process alternates. The galvo mirror, in cooperation with the SLM, exposes the corresponding fringes in their respective numbered sub-ROIs, completing a frame containing six original sub-ROIs directly.

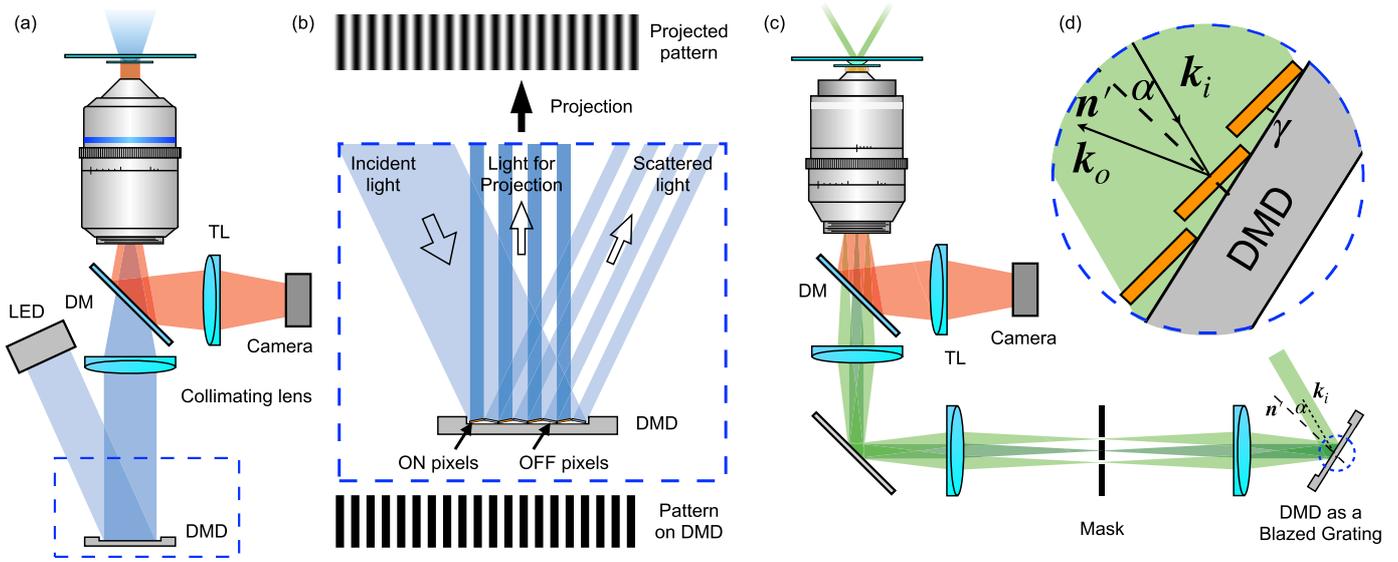

**Fig. 6 DMD-based SIM strategies.** (a) Overview of DMD-SIM with incoherent LED light source. The incoherent light from LED is shown in blue. (b) Illustration of details of DMD modulation in the dashed box in (a). As the incident light beam is incoherent, the light beam will mainly be reflected on DMD following intuitive geometric laws. The pattern on DMD is programmed as stripes, and the ON and OFF pixels reflect light beams to different directions. The modulated light with stripe patterns is collected by the collimating lens and then projected on the sample plane. Due to the low-pass filtering effect of practical optical systems, the projected pattern will be blurred with a sinusoidal like pattern. (c) Overview of laser interference-based DMD-SIM (LiDMD-SIM). The coherent laser beam, as illustrated in the above figures, is shown in green. DMD works as a blazed grating that diffracts the laser beam to -1, 0, and +1 orders. For 2D-SIM, a mask blocks the 0-order beam and allows the ±1-order beams to pass and finally interfere at the sample plane, as shown in light green. For 3D-SIM, the 0-order beam also passes through the mask, as shown in bottle green. (d) Illustration of details of DMD modulation in the dashed circle in (c). The DMD works as a blazed grating with a blazing angle $\gamma$. The normal vector of a single micromirror is noted as $n'$ and the wave vectors of incident and outgoing beams are noted as $k_i$ and $k_o$, respectively. When the angle between the normal direction and the incident beam equals to a specific value $\alpha$, i.e., $\langle n', -k_i \rangle = \alpha$, the DMD can be seen as a blazed grating. DM: dichroic mirror; TL: tube lens.

### Galvanometers

Effective generation and control of interference fringes can also be achieved through precise manipulation of two laser beams, where galvanometers play a key role. Chung[77] and Brunstein[78] attempted to achieve phase changes of two beams using mirrors mounted on piezoelectric stages, but the acquisition speed was much lower than the systems with SLMs and DMDs. With the advent of scanning galvanometers, a family of high-speed devices with excellent light modulation ability, galvo-based SIM gradually gained popularity. These SIM systems typically feature two fully symmetric scanning units, where galvo mirrors control the scanning path and frequency of the light, while the phase is controlled by a mirror fixed on a piezoelectric platform in one of the paths. In 2018, Chen *et al.* developed a TIRF-SIM system using galvos with kHz level scanning frequencies, and successfully captured the oscillatory behavior of microtubules in live cells.[79] The optical setup is presented in Fig. 7, while the polarization modulators are omitted for simplification. In 2020, Roth *et al.* proposed a TIRF-SIM system based on a Michelson interferometer and achieved simultaneous dual-color imaging.[22] In 2022, Yuan *et al.* modulated dual-color structured light using a galvo system, enabling simultaneous imaging of microtubules and actin in living cells and observing the dynamic movement of mitochondria and microtubules.[80] In the same year, Xu *et al.* replaced the piezoelectric platform with an electro-optical modulator and built a versatile system with a scanning frequency greater than 2 kHz, capable of switching between 2D-SIM, 3D-SIM, and TIRF-SIM modes.[81] This system achieved an imaging rate of 151 Hz over a 100 pixel×512 pixel field of view, with a maximum real-time reconstruction speed of 25 fps.

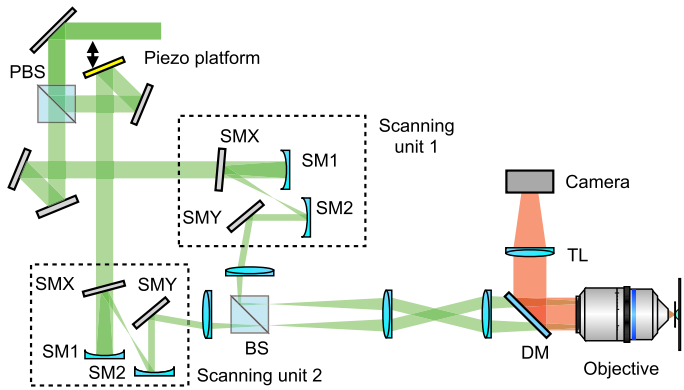

**Fig. 7 Optical scheme of a galvo-based SIM system.** Scanning is conducted by two symmetric scanning units, and phase shift is carried out by piezo platforms. PBS, polarizing beam splitter cube; BS, beam splitter cube; M1, SM1, SM2, spherical mirrors; SMX, SMY, scanning galvo mirror; TL, tube lens; DM, dichroic mirror.

As a brief conclusion, for wide-field SIM based on interference or pattern projection, high-speed devices like SLMs, DMDs, and galvos have greatly pushed forward SIM techniques towards higher spatiotemporal resolution. Details of system parameters have been listed in Table 1.

## Point-scanning based SIM

In point scanning-SIM, out-of-focus background fluorescence is effectively rejected, and point scanning-SIM techniques show inherent advantages in volumetric imaging with high contrast. To carry out point scanning, high speed devices for laser scanning are applied in point scanning-SIM implementations, including DMDs, galvos, and spinning disks.

### Digital micromirror device

As we mentioned before, although ISM improves the signal strength, there is a sharp increase in the data volume and acquisition time. To address this, York *et al.* proposed multifocal SIM (MSIM) in 2012 (Fig. 8a), where a DMD is used to generate sparse 2D excitation patterns, and image reconstruction is performed through digital relocation and deconvolution techniques, achieving a temporal resolution of 1 Hz.[49] Another advantage of DMD in MSIM lies in the accurate and fast step-wise scan control. With the MSIM system, they observed the interactions of myosin in cells.

### Galvanometers

To reduce the volume of raw data, Roth[82] and De Luca[83] independently proposed the concept of optical reassignment in 2013. They developed Optical Photon ReAssignment (OPRA) microscopy (Fig. 8b) and rescan confocal microscopy (RCM) systems, respectively. The key of both techniques lies in rescanning: the fluorescence emission is scanned by galvo for two times. In OPRA, the fluorescence beams are enlarged through lens groups for smaller focus size, while in RCM, another galvo rescans the fluorescence beam with a larger scanning amplitude, leading to the same result. Both techniques perform the photon reassignment *optically*, which avoids the need for data processing.

In the same year, York incorporated the concept of optical reassignment into his system, proposing instantaneous SIM (iSIM, instant SIM).[84] This technique increased data acquisition speed by 2-3 orders-of-magnitude. The key to the speed improvement in iSIM is that it carries out each *digital* photon reassignment step in MSIM through *optical* means (Fig. 8c). The data acquisition and processing in MSIM can be summarized as seven steps: I) Sparse multifocal illumination on the sample. II) Recording fluorescence images with a camera. III) Excluding out-of-focus light through *digital pinholing*. IV) Contracting the fluorescence image of each digital pinhole by a factor of two. V) Repeating the above steps at different positions of the multifocal excitation pattern until the entire field of view is illuminated. VI) Digitally summing the resulting images for an SR image. VII) Deconvolution for two-fold resolution enhancement. However, steps (I)–(VI) in iSIM are entirely done with microlens arrays (Fig. 8c: i, iii), matched physical pinholes (Fig. 8c: ii), and a galvanometer (Fig. 8c: iv). In another word, they are performed with optical elements purely in the hardware system, directly attaining images with resolution improvement for deconvolution in step VII. iSIM maintains a lateral resolution of 145 nm and an axial resolution of 350 nm at a sampling speed of up to 100 Hz. It is a typical example for the combination of computational and physical methods for optical sectioning. It enables 3D SR imaging of samples thicker than 20 μm with easy sample preparation and extremely high imaging speed. Currently, the main limitation on system speed comes from the inertia of the sample stage. Under strictly definitions, the high-speed devices in iSIM are the galvo and the sCMOS camera. However, comparing MSIM and iSIM, the employment of microlens arrays, physical pinhole arrays, and other optical elements in the iSIM system significantly accelerates image acquisition and processing likewise. Therefore, they can also be thought of as high-speed devices under more general definition.

The principle of Airyscan has been introduced before. A classic implementation of Airyscan technology is embodied in the LSM 980 microscope with Airyscan 2, developed by Zeiss.[85] In LSM 980, the scanning element is implemented through two independent galvanometric scanning mirrors. These mirrors are responsible for directing the laser beam across the sample in a raster pattern, enabling precise point-by-point illumination of the specimen. The unique detection array of Airyscan consists of 32 circularly arranged detector elements, each functioning as a 0.2 AU confocal pinhole, providing superior spatial resolution. Together, these elements operate equivalently to a pinhole with a higher AU value of 1.25, achieving remarkable light efficiency and an enhanced SNR. What's more, the LSM 980 system provides an imaging speed of over 25 fps in different modes. Its efficient optical design allows simultaneous acquisition of up to 36 spectral channels providing exceptional spectral flexibility and supporting complex multicolor experiments. With the Airyscan Joint Deconvolution (jDCV) for post-processing, LSM 980 achieves a lateral resolution of 90 nm and an axial resolution of 270 nm. This represents a great improvement in resolution compared to wide-field imaging and SNR improvement compared to traditional LSCM using a 1 AU pinhole.

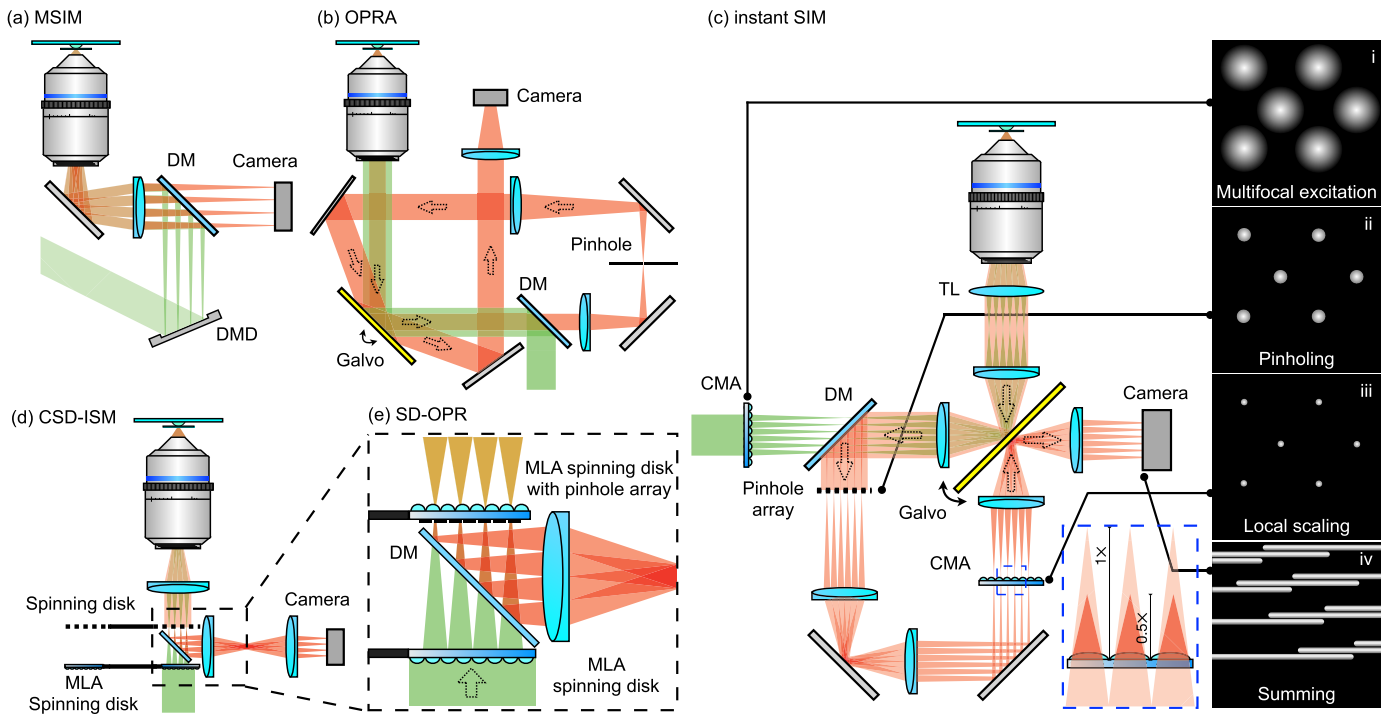

**Fig. 8 Scanning SIM strategies.** (a) Multifocal structured illumination microscopy, MSIM. A DMD generates divergent light beams that are parallel to each other. The tube lens ensures the light beams entering the objective's BFP are collimated in order to focus on the sample plane for multifocal excitation. A camera is used for multi-point detection. (b) Optical photon reassignment microscopy, OPRA. In the figure, the propagation directions of light beams are labelled with arrows, the smaller dotted ones indicating the emission fluorescence and the only large one for the excitation beam. The galvo scans the fluorescence emission two times, which are termed as descanning and rescanning, respectively. (c) instantaneous structured illumination microscopy, instant SIM (iSIM). The converging microlens array (CMA) generates convergent light beams for multifocal excitation (i). The galvo scans the excitation beams and descans the emission fluorescence. The pinhole array blocks out-of-focus fluorescence (ii). The emission beams are converged to a plane close to the second CMA, and their sizes are reduced to half of their original size (dashed box). The fluorescence beams are summed on the camera through rescanning by the galvo, and a super-resolution image is generated. In this figure, excitation beam is shown in green, and the fluorescence emission in red. DM: dichroic mirror. CMA: converging microlens array. TL: tube lens. MLA: microlens array. (d) Confocal spinning disk image scanning microscopy, CSD-ISM. In the system there are two synchronized spinning disks of the same specification. The lower spinning disk is equipped with a microlens array (MLA), which converges the incident light beams and the focal plane falls right on the upper spinning disk, which is a pinhole array. Typically, the focal length of MLA is rather small, which means the dichroic mirror (DM) between the two spinning disks should be smaller in size. Therefore, the efficient clear aperture of the CSD-ISM system is smaller. (e) Optical reassignment with spinning disks (SD-OPR). The basic optical setup is similar to that of CSD-ISM, while it replaces the spinning disk in CSD-ISM near the objective with an MLA spinning disk with a pinhole array, whose working principle is the same as that shown in the blue dashed box in (c).

In 2019, Castello *et al.* introduced a novel implementation of ISM based on a single-photon avalanche diode (SPAD) array, enabling super-resolution fluorescence lifetime imaging (FLIM).[86] The approach transforms conventional confocal microscopes into high-resolution systems by integrating a SPAD module and utilizing adaptive pixel reassignment (APR) for superior resolution and SNR. It supports multicolor, live-cell, and deep imaging without pre-calibration. The method achieves significant resolution and precision enhancements over traditional confocal and FLIM systems while reducing phototoxicity. This versatile platform represents a breakthrough for advanced biomedical imaging and time-resolved spectroscopy. In 2023, Zunino *et al.* further designed and developed an open-source tool named BrightEyes-ISM for ISM dataset image reconstruction and quality assessment.[87] In the same year, Shen *et al.* used two synchronized galvo mirrors for image rescan, proposing confocal rescan SIM (CR-SIM).[88] This method achieves super-resolution in two orthogonal directions without rotating the illumination pattern.

The relatively low imaging speed makes it difficult for ISM to be applied in *in vivo* biological imaging. In 2024, Ren *et al.* successfully combined confocal image scanning with SIM, proposing the multi-confocal ISM (MC-ISM), which strikes a good balance between pinhole diameter and pitch and well eliminates the out-of-focus signals. Instead of using active devices such as DMD or SLM, a passive pinhole array is employed in MC-ISM, followed with scanning through galvos.[89] MC-ISM increases the imaging speed by 16 times compared with MSIM, enabling deep tissue super-resolution imaging.

### *Spinning disk*

Considering the poor light utilization efficiency of DMDs, Schulz *et al.* in 2013 implemented multifocal excitation without altering the optical or mechanical components of the spinning disk confocal microscope.[90] They developed the confocal spinning disk ISM (CSD-ISM) system (Fig. 8d), achieving a temporal resolution of 3.3 Hz, although CSD-ISM still required an acquisition of 250 images.

The application of microlens arrays with a difference in the NA of excitation and detection generates optical super-resolution. The spinning disk optical reassignment (SD-OPR) is a great example.[91] As shown in Fig. 8e, in this system, a second layer of microlens array is seated below the pinhole array. In such configuration, the excitation NA is kept unchanged, while the detection NA is doubled optically. A 2× relay optics is installed to host the large NA for detection. Therefore, an SR final image is obtained optically. Benefiting from the high scanning speed of spinning disks, the system features a resolution of 163 nm laterally and 405 nm axially, with an imaging speed of up to 2000 fps.

To seamlessly integrate new technologies with commercial equipment, Hayashi *et al.* in 2015 upgraded a commercial spinning disk microscope to a spinning disk SRM by replacing the commercial disk, capturing the fusion and fission of the outer mitochondrial membrane at 100 Hz.[92] In 2021, Qin *et al.* provided a detailed tutorial for converting a commercial CSD microscope into a CSD-ISM microscope.[93] Combining deconvolution techniques, they achieved a two-fold resolution enhancement, enabling researchers in optics, microscopy, and related fields to upgrade their commercial CSD systems in about three days.

The key parameters of different point scanning-SIM techniques are also listed in Table 1.

## Applications

Traditional fluorescence microscopy plays a crucial role in the field of life sciences due to its strong contrast and specific recognition of labels. However, due to the diffraction limit, these microscopes have a limited spatial resolution, which restricts our observation of smaller and more delicate biological structures. As a super-resolution technique, SIM breaks through the limitations of the diffraction limit, doubling the resolution compared to traditional methods, allowing observers to capture finer biological details. Similarly, by combining the specificity of fluorescence labeling and the rapid response of structured light modulators, SIM stands out in the field of biological research with its high spatiotemporal resolution and low phototoxicity, making it highly favored. In the following sections, we will delve into how SIM significantly enhances the quality and depth of life science research from multiple dimensions, including resolution, imaging speed, multicolor imaging capabilities, and sample penetration depth.

### Clearer structural features

Cells, as the fundamental units of life activities, are minuscule yet exquisite systems. Structured illumination microscopy, with its super-resolution imaging capabilities that surpass the diffraction limit, provides tools for the study of cellular and subcellular structures, which is of significant importance for revealing the essence of life phenomena and the mechanisms underlying certain diseases.

To illustrate how SIM can be applied in the study of life science, we can take the interaction between cells and viruses as an example. Viruses are small RNA-based organisms that rely on host cells for survival. They infect living cells, forcing them to synthesize viral proteins, ultimately causing damage to the host[94]. High resolution SIM plays an important role in investigating cell-virus interactions[95-97], and it helps to explore pathogenic mechanisms. The John Cunningham virus (JCV) is a kind of virus associated with progressive multifocal leukoencephalopathy (PML), targeting the promyelocytic leukemia nuclear bodies (PML-NBs) within the cell nucleus. Shishido-Hara *et al.* used fluorescence labeling of PML protein and JCV capsid protein, and observed the localization of JCV in PML-NBs using SIM technology.[95] As shown in Fig. 9a, the PML signal is intense at the outer shell of larger PML-NBs, forming a ring-like structure, suggesting that the virus may proliferate extensively in these areas. This study provides important clues for understanding the pathogenic mechanism of JCV.

The resolution of SIM can be significantly improved with nonlinear SIM, enabling the resolution of finer biological

**Table 1. Key parameters of SIM techniques**

| Technique | Key element(s) | Lateral resolution (nm) | Axial resolution (nm) | Imaging speed level* | Imaging depth level** | Reference |
|---|---|---|---|---|---|---|
| INTERFERENCE | | | | | | |
| 2D-SIM | Grating | 115 | / | ★ | ★★★ | Gustafsson, 2000 |
| 3D-SIM | Grating | 104 | 279 | ★ | ★★ | Gustafsson, 2008 |
| N-SIM E | Grating | 115 | 269 | ★ | ★ | NIKON, 2016 |
| TIRF-SIM | SLM | ~100 | / | ★★ | ★ | Fiolka, 2008; Kner, 2009 |
| 2D-SIM | SLM | 144 | / | ★ | ★★★ | Chang, 2009 |
| 3D-SIM | SLM | 120 | 360 | ★ | ★★ | Shao, 2011 |
| Dual color 3D-SIM | SLM | 110 | 360 | ★ | ★★ | Fiolka, 2012 |
| Fast SIM | SLM | ~100 | / | ★★★ | ★ | Förster, 2014; Lu-Walther, 2015; Song, 2016 |
| Four-beam SIM | SLM | 123.5 | 163 | ★ | ★★ | Li, 2023 |
| Hessian TIRF-SIM | SLM | 88 | / | ★★★★★ | ★ | Huang, 2018 |
| PAR-SIM | SLM, Galvo | 100 | / | ★★★★★★★ | ★★★ | Xu, 2024 |
| 2D DMD-SIM | DMD | ~130 | / | ★★★ | ★★★ | Li, 2020; Lachetta, 2021; Gong, 2023 |
| DMD-SIM | DMD | 133 | 300 | ★ | ★★ | Li, 2024 |
| TIRF-SIM | Piezo stage | 100 | / | ★★★★ | ★ | Brunstein, 2013 |
| TIRF-SIM | Galvo | 105 | / | ★★★ | ★ | Chen, 2018; Roth, 2020 |
| Multi-mode SIM | Galvo | 90 | ~300 | ★★★★★ | ★ | Xu, 2022 |
| PROJECTION | | | | | | |
| DMD-LED SIM | DMD | 90 | 930 | ★★★ | ★★★★★ | Dan, 2013 |
| POINT-SCANNING | | | | | | |
| ISM | Piezo scanning mirror | 150 | ~350 | ★ | ★★★★ | Müller, 2010 |
| MSIM | DMD | 145 | ~350 | ★★ | ★★★ | York, 2012 |
| OPRA | Galvo | ~140 | ~450 | ★★ | ★★★★ | Roth, 2013 |
| RCM | Galvo | ~140 | ~450 | ★★ | ★★★★ | De Luca, 2013 |
| iSIM | Galvo | 145 | 350 | ★★★★ | ★★★ | York, 2013 |
| LSM 980 Airyscan (jDCV) | Galvo | 90 | 270 | ★★ | ★★★★ | ZEISS, 2020 |
| LSM 980 Airyscan | Galvo | 120 | 350 | ★★★ | ★★★★ | ZEISS, 2020 |
| CSD-ISM | Spinning disk | 130 | ~400 | ★★ | ★★★ | Schulz, 2013 |
| SD-OPR | Spinning disk | 105 | 267 | ★★★★★ | / | Azuma, 2015 |
| SD-SRM | Spinning disk | 120 | / | ★★★★ | / | Hayashi, 2015 |
| CR-ISM | Galvo | 133 | / | ★★★ | ★★★★★★ | Shen, 2023 |
| MC-ISM | Galvo | 130 | 330 | ★★★ | ★★★★★★ | Ren, 2024 |
| AO 2P-MSIM | SLM | 142 | 493 | ★★ | ★★★★★★ | Zhang, 2023 |
| NSPARC | SLM | 100 | 300 | ★★ | ★★★★★★ | NIKON, 2020 |

*The imaging speed is normalized and divided into seven levels. ★ <1 fps, 1 fps < ★★ < 10 fps, 10 fps < ★★★ < 50 fps, 50 fps < ★★★★ < 100 fps, 100 fps < ★★★★★ < 180 fps, 180 fps < ★★★★★★ < 220 fps, ★★★★★★★ > 220 fps.

**The imaging depth is divided into seven levels. ★ <5 μm, 5 μm < ★★ < 10 μm, 10 μm < ★★★ < 50μm, 50 μm < ★★★★ < 110 μm, 110 μm < ★★★★★ < 170 μm, ★★★★★★ > 170 μm.

structures. Rego *et al*. combined the nonlinear response of the reversible photo-switchable protein Dronpa to generate higher-order harmonics (HOH), achieving a resolution of 42 nm, four times that of conventional microscopes.[26] They used this technology to image the nuclear pore complex, and the imaging results are shown in Fig. 9b, revealing that two nuclear pore proteins (Nup98 and POM121) have different localization patterns: POM121-Dronpa is a complete membrane protein with many rings having an inner diameter of 40-70 nm, while Dronpa-Nup98 appears as uniform small punctate dots. This is the first time the ring structure of the nuclear pore complex has been resolved in mammals, and this breakthrough provides important data for a deeper understanding of the structures within the cell nucleus.

Li *et al*. used high-NA objectives and increased the resolution to 84 nm, successfully observing more complex cellular structures—actin filaments.[98] They imaged clathrin-coated pits (CCPs) in COS-7 cells using fluorescence labeling and found ring-like structures shown in Fig. 9c. The size of these rings was similar to that of the CCPs, suggesting that they may be related to other clathrin-independent endocytosis. By comparing the distribution of the two fluorescent proteins, the role of the actin ring in the formation and function of CCPs can be further explored. Therefore, the improvement in resolution is crucial for understanding the relative distribution of CCPs and the cytoskeleton.

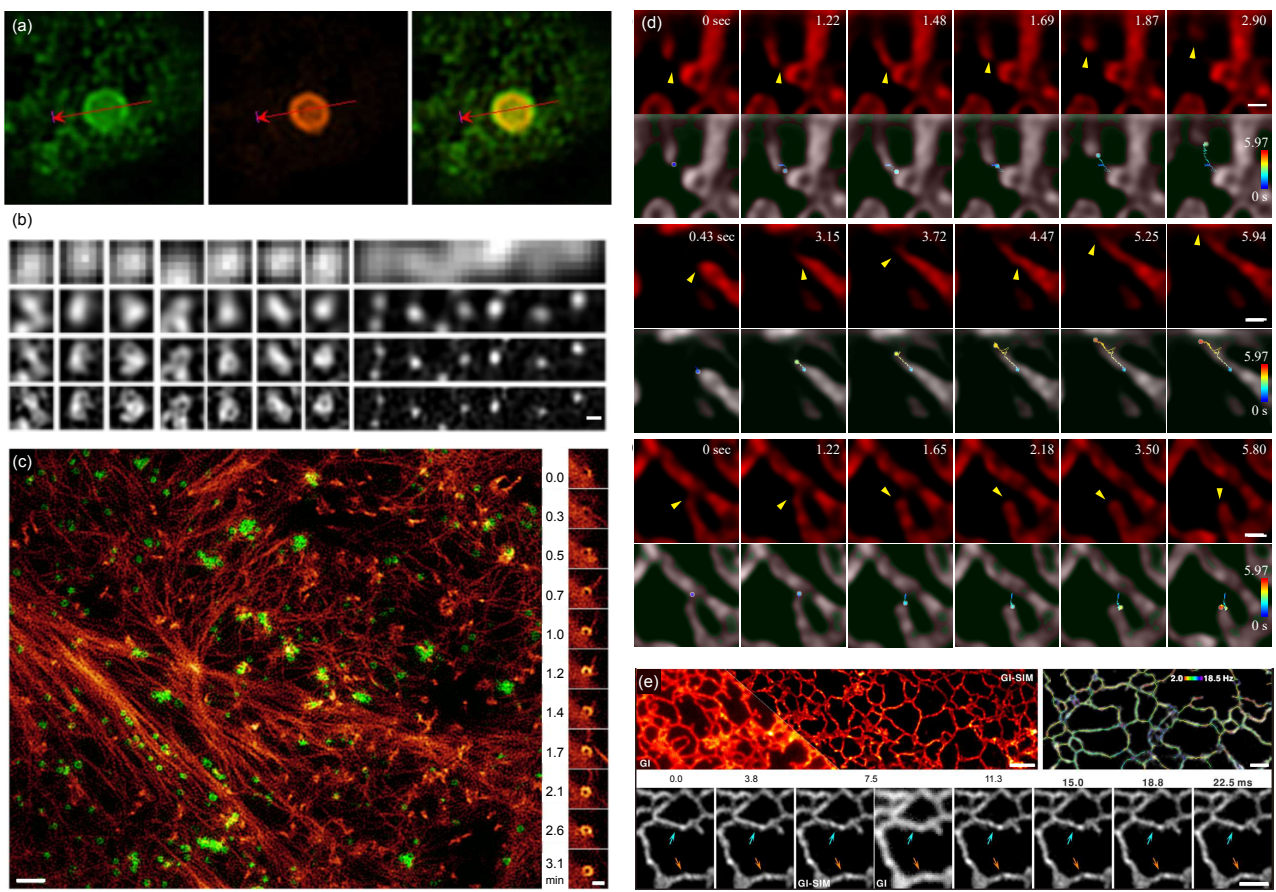

**Fig. 9 SIM reveals subcellular structures and dynamic processes with high spatiotemporal resolution.** (a) Precise PML-NB organization in JC virus-infected human glial cells detected by SIM. JC virus VP1 labeled with Alexa Fluor 488 (green), PML protein labeled with Alexa Fluor 568 (red). VP1 proteins encircle the outer surface of the PML-NB shell. The peak distance of VP1 protein (1.2 μm) was slightly longer than that of PML protein (1.0 μm).[95] (b) SIM image of nuclear pores in the nuclei of mammalian cells. Human HEK293 cells were transiently transfected with Dronpa-Nup98 or POM121-Dronpa. Left inset shows POM121-Dronpa, right shows Dronpa-Nup98 imaging. Rows from top to bottom show TIRF, linear SIM, NL-SIM with one HOH, and NL-SIM with two HOHs. Scale bar: 100 nm.[26] (c) COS7 cell image showing individual CCP and clathrin-coated pits (green) and cortical f-actin (red) at a single time point, with the formation of an f-actin nanoscale ring. Scale bar: 1 mm (left), 200 nm (right).[98] (d) PAR-SIM captures mitochondrial kiss-and-run and extrusion events, with time-encoded pseudo-color trajectories. Timestamps appear in the top right of each sub-image. Scale bar: 0.5 μm.[70] (e) GI-SIM image of the ER network in live COS-7 cells expressing mEmerald-KDEL captured at 266 fps. Scale bar: 2 μm. The upper right shows ER tubule skeleton with color-coded oscillation frequencies overlaid. The bottom shows time-lapse images of ER tubule dynamics. Cyan and orange arrows indicate formation and disappearance of constriction sites. Scale bar: 1 μm.[107]

In addition to improvements in optical mechanical components, algorithms also play a crucial role in the resolution enhancement of SIM. Zhao *et al.* utilized prior knowledge of the sparsity and continuity of biological structures to propose Sparse-SIM, which produced excellent restoration effects for the imaging of vesicles under wide-field illumination.[99] This technology can resolve fusion pores in INS-1 cells, which are smaller than the pores detectable by conventional TIRF-SIM, and can even resolve fusion pores with a fitted diameter of 61 nm. With multi-resolution analysis (MRA), Hou *et al.* improved the resolution of SIM to around 60 nm.[21] Fine actin filaments and the ER network can be clearly resolved.

The continuous development of SIM technology not only enhances the resolution of complex biological structures but also opens new avenues for in-depth study of viral infection mechanisms and cellular functions. In the future, with further optimization of the technology and expansion of its application range, SIM will provide more possibilities for basic research in life sciences and clinical diagnostics.

**Faster observation of dynamic processes**

Although techniques like stochastic optical reconstruction microscopy (STORM) achieve higher resolution than SIM, they present limitations in observing live cells and dynamic processes.[100] In contrast, SIM typically requires the acquisition of 9-15 raw images for super-resolution reconstruction, making it ideal for capturing rapid dynamics.

Mitochondrial morphology and crista damage are closely related to various diseases, including cancer and osteoarthritis[101-103]. Given the complexity, heterogeneity, and highly dynamic nature of mitochondrial structure[104], super-resolution techniques are essential to observe their dynamic behaviors. Using SIM, researchers have observed mitochondrial fragmentation, donut formation[105], and contact interactions[106]. However, spatial and temporal resolution limitations hinder inner membrane observation. Hessian-SIM[69], with an 88 nm spatial resolution and a temporal resolution of 188 Hz using low-photon doses, provides clear visualization of mitochondrial cristae dynamics in live cells. This technique enables real-time observation of crista remodeling during mitochondrial fission and fusion processes. To achieve higher imaging speeds, Xu *et al.* proposed PAR-SIM through hardware improvements. Utilizing PAR-SIM technology, they realized ultra-fast imaging of mitochondrial dynamics with a temporal resolution of 408 Hz.[70] This allowed for comprehensive capture of mitochondrial tubulation and fusion, including the rapid formation of mitochondrial dynamic tubules (MDT) and the extrusion and kiss-and-run interactions among mitochondria, as illustrated in Fig. 9d. Time-encoded pseudo-color trajectories detail MDT's rapid movement. Statistical analysis of MDT displacement and velocity reveals a maximum MDT extension rate of 21.24 μm/s at 136 Hz with PAR-SIM, while conventional SIM at 5.44 Hz only captured a peak velocity of 3.82 μm/s. Such insights are key to understanding mitochondria's role in cellular function and pathology.

The endoplasmic reticulum, with its intricate network structure, undergoes rapid remodeling based on cellular environmental shifts, demanding imaging technologies with high spatiotemporal resolution. Guo *et al.*[107] used GI-SIM to capture the oscillatory behavior of the ER's complex reticular structure and tubules at 266 fps. They used color coding to represent the oscillation rate of ER tubules, finding the highest oscillation frequencies in denser tubules. Additionally, they observed ER tubule contraction within 10 ms, speculating that these changes may be related to ER functional regulation during cellular stress or metabolic activities, as shown in Fig. 9e. This high spatiotemporal resolution technique offers a valuable tool for elucidating the dynamics of the ER, demonstrating its capability for rapid imaging of complex biological phenomena.

Rapid imaging of SIM results in shorter exposure times, leading to a lower SNR, which poses a critical challenge in super-resolution reconstruction. To address this limitation, Qiao *et al.* introduced a rationalized deep learning (rDL) approach, which leverages prior knowledge of illumination patterns to denoise captured images and enhance high-frequency details.[108,109] By applying the rDL GI-SIM technique, the researchers achieved a remarkable spatial resolution of 97 nm and a temporal resolution of 684 fps, enabling the clear visualization of the hollow tubular structure of the ciliary axoneme and the characteristic back-and-forth motion of a cilium. Furthermore, with two-color rDL GI-SIM imaging, they successfully captured the intricate behavior of intraflagellar transport (IFT) trains during spatio-temporal remodeling.

High-temporal resolution SIM offers a powerful tool for understanding subcellular dynamics and their roles in disease, benefiting foundational life science research and disease study.

**Richer insights into organelle interactions**

As an integrated entity, cells require examination not only of the morphology and dynamics of individual organelles but also of the interactions between them.

Cytoskeleton is crucial to the realization of cell division, growth and morphogenesis and other basic functions. Due to its vast extension area, it can contact and interact with a variety of organelles. Fiolka *et al.* labeled clathrin and actin in HeLa cells to investigate the interactions between the cytoskeleton and vesicular systems, employing two-color 3D-SIM to detail actin filament and clathrin-coated vesicle dynamics.[63] Observations revealed clathrin-coated vesicles as highly dynamic dots or rings that emerge, dissipate, and disassemble within the cell, with precise tracking of their division. This is pivotal for understanding the role of clathrin-coated vesicles in cellular transport and membrane dynamics.

Guo et al.[107] provided a fresh perspective on membrane contacts and functional interactions between organelles using GI-SIM, unveiling various inter-organelle dynamic interactions. For the first time in mammalian cells, they confirmed the *hitchhiking* phenomenon, where organelles attach to moving organelles through membrane contact sites for co-transport. Fig. 10a shows ER tubules attached to mitochondria, extending as the mitochondria move, while the lower image illustrates richer mitochondrial interactions with late endosomes (LE) or lysosomes (Lyso). When LE or Lyso approaches mitochondria, significant changes in mitochondrial morphology occur, sometimes forming tubular bridges between mitochondria, resulting in complete fusion. This phenomenon supports mechanisms underlying mitochondrial DNA integrity and material exchange.

Moreover, Gong et al. reported a four-color SIM imaging technique, enabling simultaneous labeling of microtubules, actin, mitochondria, and nuclei for four-channel fluorescence imaging in the same field of view.[75] Fig. 10b shows fluorescence images of multi-channel cell structures and enlarged views of single structures. Time-series imaging resolved the dynamics of various cellular structures, such as mitochondrial morphology and microtubule oscillation.

Multi-color imaging of SIM provides unique insights into the complex interactions between organelles. For example, studies have revealed interactions between mitochondria and microtubules[110,111], mitochondria and lysosomes[112,113], mitochondria and ER[114], ER and vesicles[111] and microtubules and actin[115,116]. SIM deepens our understanding of organelle functions and interactions in cellular regulation, providing powerful tools to explore disease mechanisms at the molecular level, particularly for pathologies involving organelle interaction anomalies.

### Higher depth three-dimensional observation

In modern life science research, multidimensional imaging is essential to capture the full spatiotemporal distribution of gene and protein expression and their dynamic developmental processes. To accurately visualize these events, long-term imaging of specific tissues or organs is required at cellular resolution[117,118].

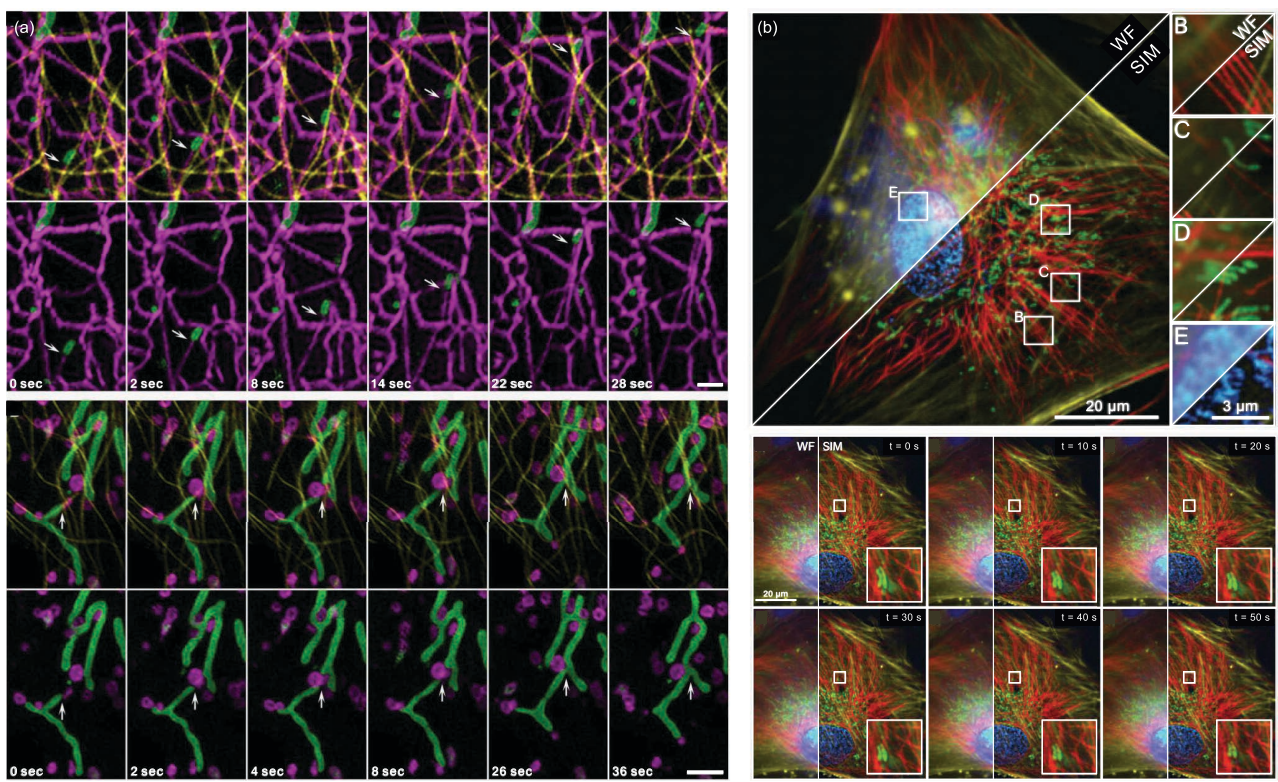

**Fig. 10 Multicolor imaging reveals more interactions between organelles.** (a) Top: Visualization of newly formed ER tubules (magenta) originating on a mitochondrion (green) and moving along a microtubule (yellow), with the tip highlighted by a white arrow. Scale bar: 1 μm. Bottom: LE- or lysosome-mediated mitochondrial (green) fusion events, with mitochondrion tips marked by white arrows. Scale bar: 2 μm.[107] (b) In live BJ fibroblast cells, the main panel shows WF and SIM images of microtubules (red), actin (yellow), mitochondria (green), and nuclei (blue), with enlarged images of regions of interest. And time-lapse images with additional inset views of microtubule fibers and mitochondrial movement.[75] Reprinted with permission from ref 75 © Optical Society of America.

Fiolka *et al.* applied 3D-SIM to image 8.125 μm-thick HeLa cells, using mCherry to label microtubules and MitoTracker for mitochondria.[65] 3D-SIM significantly improved resolution and effectively suppressed out-of-focus light, allowing mitochondrial cristae to be observed under an optical microscope, as shown in Fig. 11a – previously only achievable via electron microscopy (EM). Dan *et al.* used 3D optical sectioning modalities to image mouse neuron cells with a thickness of 74 μm and pollen grains with a thickness of 120.8 μm.[16] The original images were then reconstructed in 3D and demonstrated the maximum intensity projection effect, as shown in Fig. 11b.

Point-scanning super-resolution techniques have also shown excellent axial imaging capability and defocus suppression. While mitochondrial inner membrane imaging in animal cells has advanced widely, plant cells present challenges due to strong scattering, autofluorescence, and cell wall structures. Ren *et al.* proposed MC-ISM and achieved *in vivo* imaging of mitochondria at a depth of 20 μm in the hypocotyl of *Arabidopsis thaliana*, unveiling for the first time the internal membrane structure of plant mitochondria. Additionally, they identified morphological similarities between plant and animal cell mitochondria, including spherical and elongated forms, and successfully documented the dynamic behaviors of these two mitochondrial morphologies, as shown in Fig. 11c.[89] Three-dimensional imaging of a 12 μm thick mouse kidney slice at 100 magnification clearly revealed separated structures, including filamentous actin, nephron components, and nuclei, offering unprecedented clarity for histological research. Additionally, MC-ISM technology is compatible with low-magnification objectives, as shown in a deep imaging example of zebrafish head tissue with 2.5 μm axial spacing and the total imaging depth is 175 μm, which highlighted fine cellular structures in Fig. 11d. This study not only provides critical visual evidence for elucidating mitochondrial function and dynamics in plant cells but also highlights the remarkable capability of point-scanning super-resolution microscopy for imaging within thick tissue samples.

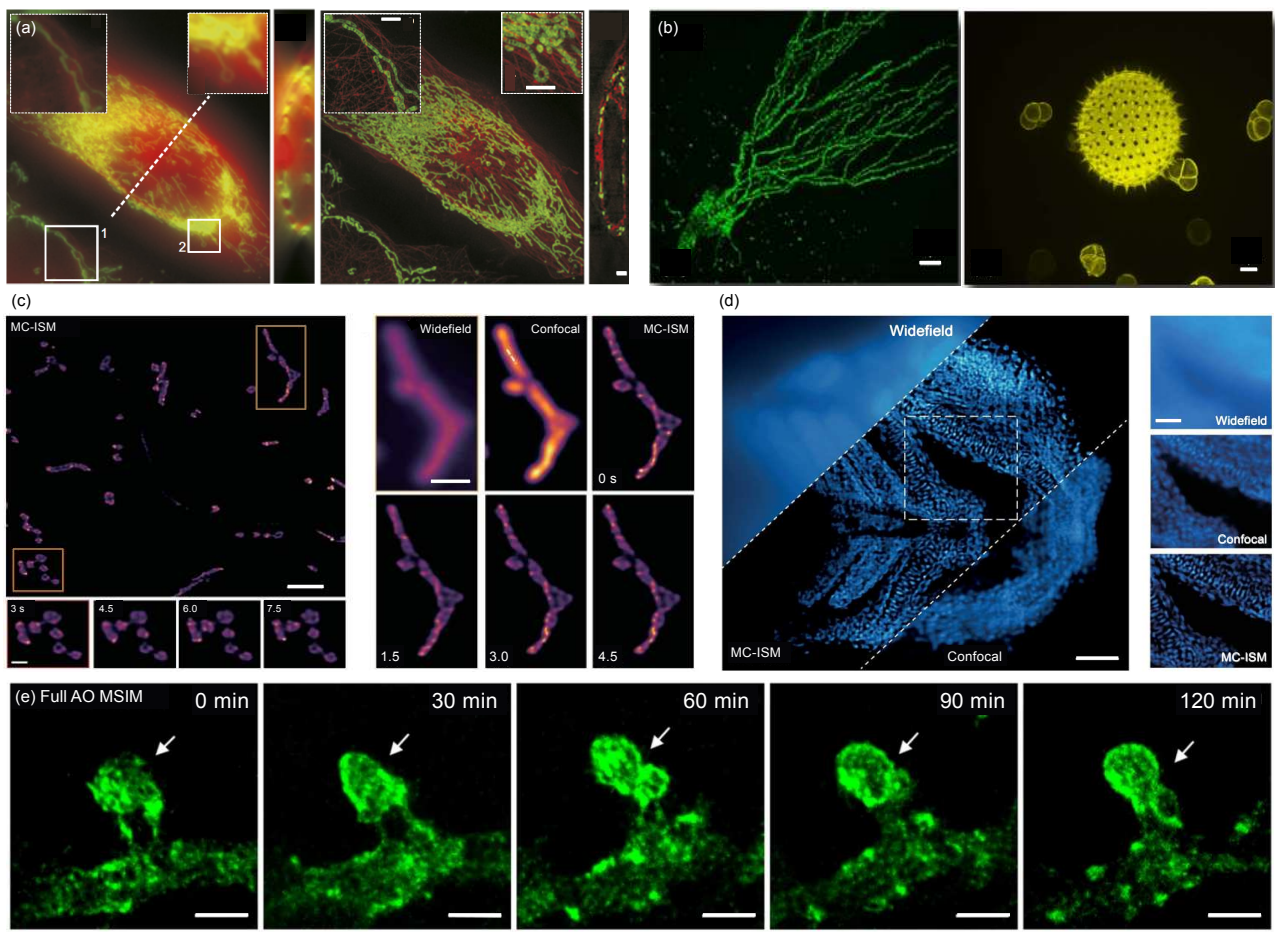

**Fig. 11 3D imaging of stripe-based SIM or point-scanning SIM enables the observation of deeper biological phenomena with enhanced imaging depth.** (a) Comparison between conventional wide-field microscopy and 3D SIM imaging. The main image is a maximum intensity projection along the z-axis, with mitochondria in green and microtubules in red. Enlarged views of Regions 1 and 2 are shown at the top left and top right, respectively. The right side of the main image shows a y-z cross-section along the dashed line. Scale bar: 2 μm.[65] (b) Optical section images of large volumes of mouse neuronal cells and mixed pollen grains with maximum intensity projection along the z-axis.[16] (c) Main image shows MC-ISM super-resolution imaging of *Arabidopsis* hypocotyl mitochondria labeled with HBmito Crimson. The bottom left shows the dynamic imaging of spherical mitochondria (in red box), scale bar: 0.5 μm. The right image shows dynamic imaging of elongated mitochondrial cristae (in yellow box). Scale bar: 1 μm.[89] (d) Comparison imaging of DAPI-stained zebrafish head using wide-field, confocal, and MC-ISM with a 20× air objective. Scale bar: 70 μm. Right panel compares three imaging modes within the white dashed box. Scale bar: 25 μm.[89] (e) Time-lapse of dynamic morphological changes in a specific axonal protrusion observed using MSIM imaging. Scale bar: 2 μm.[119]

Dynamic imaging at greater depths represents a prominent advantage of point-scanning super-resolution imaging techniques. During the development of neuronal networks, axonal elongation and synapse formation are critical processes for constructing neural networks. However, due to the limitations in imaging depth, it is often challenging to observe these phenomena with high clarity. Zhang et al. utilized adaptive optical two-photon multifocal structured illumination microscopy (AO 2P-MSIM) to perform long-term imaging of motor neurons in 2-dpf-old (dpf, days post-fertilization) zebrafish larvae at a depth of 218 μm.[119] This approach not only revealed the intricate structural details of axons and protrusions in zebrafish motoneurons but also captured dynamic morphological changes in axonal protrusions, as illustrated in Fig. 11e.

These examples merely scratch the surface of 3D imaging applications in biological research. Researchers have observed various other complex biological processes, such as *Drosophila* embryo development[118], and zebrafish heart dynamics[49,120]. These observations further expand our understanding of large-scale and dynamic biological processes, offering new perspectives for studies in functional genomics and human disease models.

## Conclusion and Outlook

In recent years, SIM has become increasingly prevalent in the biomedical field, emerging as a crucial tool for studying cellular functions and structures. The high spatiotemporal resolution capabilities allow scientists to observe microscopic structures and dynamic processes within cells with unprecedented clarity, unveiling biological phenomena that elude traditional microscopes.

The increasing complexity of in-depth research necessitates more sophisticated system performance, propelling SIM towards high fidelity extended resolution 2D imaging, and 3D super-resolution imaging for thick specimen. These advancements are closely linked to high-speed super-resolution techniques, as the sample usually remains static during the imaging process; otherwise, motion artifacts may occur. Consequently, the evolution of SIM across various dimensions has stimulated the development of advanced high-speed modulation devices for broader applications. On one hand, there is a trend towards reducing mechanical movement in traditional components and adopting digital modulation devices with swift response times, such as SLMs, DMDs, LC polarization modulators, and phase retarders. This shift is crucial for fulfilling the high-speed SIM demands and leveraging SIM's capabilities in live-cell imaging. On the other hand, the direct integration of optical elements into the system for parallel imaging and optical reconstruction, which combines high parallelism and processing efficiency, represents another innovative trend. Examples include multi-focus scanning in MSIM versus single-spot scanning in ISM, or the microlens arrays in iSIM versus digital photon reassignment in MSIM. Both trends pave new ways to enhance SIM's spatial and temporal resolution. As previously mentioned, these high-speed devices play a significant role.

For comparison, a series of radar maps is presented in Fig. 12 to show the performances of the SIM techniques covered in this review, including excitation intensity (phototoxicity), imaging speed, axial resolution, lateral resolution, and imaging depth. It is necessary to mention that different microscopy techniques have their own pros and cons, each tailored to specific applications. There is no *A panacea for all ills* method that simultaneously offers high resolution, high speed, large FOV, multidimensional capabilities, multicolor imaging, multimodal adaptability, low cost, low phototoxicity, and deep imaging capabilities. On one hand, we should choose the exciting method that best suits the experiment requirements, and on the other hand, we shall keep pushing forward the development of future SIM technologies toward better performance in these aspects.

Over the past decade, there has been a growing trend in biological research to integrate SRM with various microscopy techniques, leveraging its unique capability to surpass the diffraction limit in optical imaging. Techniques such as STORM and stimulated emission depletion microscopy (STED)[6] have been combined with EM[121] and FRET[122]. However, these methods face challenges due to the stringent requirements for photoswitching fluorophores, the slow imaging speeds associated with STORM, and the photobleaching issues inherent in STED, hinder the achievement of superior image quality and further advancements.[14] In contrast, SIM stands out for its gentle approach—characterized by low phototoxicity and photobleaching—and its versatility, being compatible with most fluorescent labels. These attributes position SIM as a promising candidate for combination with other techniques, offering enhanced microscopy strategies. The synergy of SIM with other high-resolution imaging modalities is poised to significantly broaden its utility in life sciences.

SIM's continuous evolution, driven by the demand for clearer, faster, and deeper biological imaging[18], positions it as a pivotal technology in life sciences, with the potential to revolutionize our understanding of cellular mechanisms and contribute to advancements in diagnostics and therapeutics.

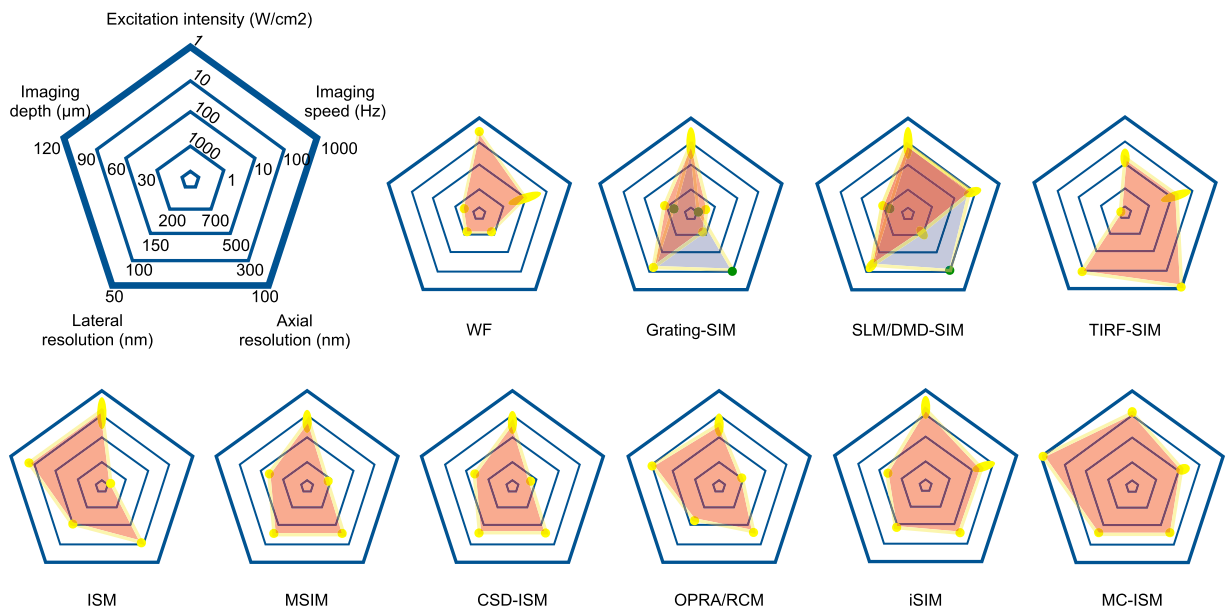

**Fig. 12 Summary and performance comparison of SIM techniques in radar maps.** The light blue area and green points stand for performances in 3D-imaging mode.

Looking ahead, we anticipate that the development of emerging and next-generation SIM will likely proceed in the following directions:

- **Multi-plane parallel imaging:** Multi-plane parallel imaging[123,124] enables the simultaneous imaging of multiple depths within a sample; cross-correlation analysis and precise plane distance adjustment further help to achieve simultaneous 3D image stack acquisition. The integration of multi-plane parallel image acquisition and 2D- or 3D-SIM ~~will be~~ is valuable in further accelerating imaging speed and improving imaging quality. For instance, image-splitting prisms and multifocus diffraction gratings have been applied in SIM for high-speed live cell imaging and contrast enhancement.[125-127] However, as the effective thickness of current structured illumination patterns is limited by the focal depth of the objective, novel imaging mechanism such as Bessel illumination should be developed to take full advantage of multi-plane imaging for SIM.
- **Metasurfaces:** Metasurfaces are cutting-edge diffraction optical elements (DOEs) made of subwavelength artificial structures.[128] Metasurfaces can manipulate the phase, polarization, and amplitude of light waves.[129-131] Therefore, they may work as key elements in generation and modulation of structured illumination. As a typical metasurface, meta-lens can achieve specific effects such as beam shaping, achromatic focusing, and wide-angle performance with careful design and material choices.[132] An achromatic meta-lens array consisting of tens of millions of nanoantennas has been applied for structured light projection in an intelligent and compact depth sensing and imaging system which can work in all light levels.[133] Alternatively, metasurface can also be used to generate the super-lattice structured illumination pattern, which can potentially improve the resolution of SIM imaging.[134] Compared with traditional DOEs, metasurfaces provide more precise light control through multi-parameter manipulation, with miniaturized system design and aberration elimination. The application of metasurfaces in SIM systems carries the potential to facilitate system miniaturization and enhance the spatial resolution, thus advancing the development of SIM.
- **Correlative microscopy:** The complementarity of EM and SRM in FOV and spatial resolution makes correlative light and electron microscopy (CLEM) very significant[135,136], especially with the development of liquid cell electron microscopy.[137,138] As SIM is less sensitive to the EM-related sample preparation, it makes the integration of SIM and EM more promising. In addition to the hardware integration of the two modalities, the development of dual contrast reagents for both EM and SIM modalities is also crucial.[139]
- **FRET-SIM:** The integration of SIM with quantitative FRET (qFRET) emphasizes precise quantification of molecular interactions at sub-diffraction scales.[140,141] SIM-FRET overcomes the challenges of traditional FRET imaging by resolving spatial details down to 120 nm while maintaining quantitative fidelity in measuring FRET efficiency $E$ and donor-to-acceptor ratios $R$. SIM-FRET avoids issues like photobleaching and signal distortion that compromise quantification. The combination not only extends FRET's resolution but also strengthens its quantitative potential, enabling researchers to analyze protein stoichiometry, binding dynamics, and functional states in real time, providing a robust platform for quantitative biological research. One major issue is the reconstruction artifacts that can arise during SIM image processing, which may distort the delicate FRET signal. Additionally, the weak FRET signals are often plagued by low SNR and strong out-of-focus background fluorescence, making accurate detection and quantification difficult. Enhanced image reconstruction methods and improved noise reduction strategies are essential for effectively combining SIM and quantitative FRET to conduct more accurate and reliable observations of molecular interactions in live cells.

- **FLIM-SIM:** FLIM measures the fluorescence lifetime of molecules, offering critical insights into the local molecular environment, such as pH, ion concentration, and protein interactions, which are essential for understanding cellular function. The integration of SIM with point-scanning[142] or wide-field FLIM[143,144] presents a transformative potential to achieve super-resolution cellular morphology imaging while simultaneously incorporating both spatial and temporal information. One major difficulty of the integration lies in the reconstruction process of SIM, which may introduce artifacts or interfere with the quantitative accuracy of fluorescence lifetime measurements. Despite the technical hurdles, the SIM-FLIM combination holds great promise. It will enable researchers to observe dynamic changes in subcellular structures with super-resolution clarity, monitor real-time protein interactions, and gain functional insights into molecular activities at an unprecedented spatial resolution, making it a vital tool for advanced live-cell imaging and molecular biology research.
- **Combination of SIM and label-free microscopy:** Label-free techniques such as 3D optical diffraction tomography (ODT) use scattered light to reconstruct 3D refractive index maps, enabling detailed visualization of cell morphology.[145,146] When integrated with label-free techniques such as 3D ODT, the super-resolution fluorescence-assisted diffraction computational tomography (SR-FACT) imaging system gains the capability to observe cellular structures at label free condition, with the advantage of SR imaging with specific fluorescence labeling.[147,148] This makes the technique ideal for long-term live-cell imaging, minimizing phototoxicity while tracking dynamic organelle interactions and complex cellular processes over extended periods.
- **Imaging in new dimensions:** The super-resolution microscopy can be further exploited in other physical dimensions, such as time, spectrum, polarization, *etc.* Polarized SIM (Polar-SIM) represents a significant advancement in SRM by offering both spatial and angular information.[149,150] Polar-SIM extends the SR imaging capability by resolving fluorescent dipole orientations within the *spatio-angular hyperspace*. This is achieved through angular harmonics reconstruction, enabling the simultaneous acquisition of high-resolution spatial images and precise polarization data. The inclusion of polarization as an additional imaging modality elevates SIM from a purely structural tool to a functional imaging platform, capable of probing the orientation, organization, and dynamics of molecular assemblies at sub-diffraction scales. Polar-SIM has been successfully applied to a range of biological systems, such as cytoskeletal networks and neuronal membrane-associated periodic skeletons, providing unparalleled insights into the angular distribution of fluorescent dipoles with a lateral resolution of ~100 nm. It also maintains quantitative fidelity across a broad range of imaging modalities, including 2D-SIM, 3D-SIM, and TIRF-SIM, ensuring compatibility with live-cell imaging and thick specimen analysis. What's more, incorporating polarization information directly into the SIM reconstruction process could help reduce imaging artifacts, thereby improving the accuracy and reliability of both spatial and angular data.[150] As a pioneering approach, Polar-SIM exemplifies how integrating functional imaging dimensions with super-resolution microscopy can address complex biological questions, particularly those involving molecular orientation and structural dynamics. In conjugation with the spectral and polarization freedom information, multiple membrane associated organelles can be distinguished simultaneously.[151] It shows us a new paradigm for advancing the understanding of cellular architecture and biophysical processes.[152]

In conclusion, the relentless pursuit of deeper biological insights has catalyzed the evolution of SIM into a formidable tool for super-resolution imaging. In this review, we categorize the SIM technologies into two: stripe-based SIM, and point-scanning based SIM. In each aspect, the quest for enhanced 2D sectioning and deeper 3D imaging has driven technological innovations, particularly in high-speed modulation devices like SLMs, DMDs, and galvanometers, which are pivotal for achieving the speed and resolution demanded by modern biological research. The integration of optical elements for parallel imaging and optical reconstruction represents a paradigm shift, offering novel strategies to boost SIM's capabilities. In the near future, its synergy with other imaging modalities promises to unravel the complexities of subcellular structures and dynamics.

**Disclosures**
The authors declare no competing financial interest.

**Acknowledgments**

This work was supported by and the National Natural Science Foundation of China (62025501, 62335008, 62405010) and the National Key R&D Program of China (2022YFC3401100).